\documentclass[a4paper,11pt]{article}
\usepackage{jcap/jcap}
\usepackage{lineno}
\usepackage{amsmath}
\usepackage{misc/macro}
\usepackage{graphicx}
\usepackage{subcaption}
\usepackage{gensymb}
\usepackage{orcidlink}
\usepackage[table,xcdraw]{xcolor}
\providecommand{\textcolour}[2]{\textcolor{#1}{#2}}
\usepackage{multirow}
\usepackage{booktabs}
\usepackage[normalem]{ulem}

\definecolor{rowgray}{gray}{0.95}

\title{The Simons Observatory: Analysis of methods for polarization calibration of multiple instruments}

\author[1]{Anto I. Lonappan\,\orcidlink{0000-0003-1200-9179}\note{Corresponding author.},}
\author[2,3]{Patricia Diego-Palazuelos\,\orcidlink{0000-0002-5129-2379},}
\author[4]{Carlos Herv\'ias-Caimapo\,\orcidlink{0000-0002-4765-3426},}
\author[5,1]{Kam Arnold\,\orcidlink{0000-0002-3407-5305},}
\author[6]{Susanna Azzoni\,\orcidlink{0000-0002-8132-4896},}
\author[7,8,9,10]{Carlo Baccigalupi\,\orcidlink{0000-0002-8211-1630},}
\author[11]{Matteo Billi\,\orcidlink{0000-0003-0616-0593},}
\author[12]{Erminia Calabrese\,\orcidlink{0000-0003-0837-0068},}
\author[13]{Yuji Chinone\,\orcidlink{0000-0002-3266-857X},}
\author[1]{Matthew Chu\,\orcidlink{0000-0001-5884-911X},}
\author[14]{Mark Devlin\,\orcidlink{0000-0002-3169-9761},}
\author[15]{Josquin Errard\,\orcidlink{0000-0002-1419-0031},}
\author[16,17]{Nicholas Galitzki\,\orcidlink{0000-0001-7225-6679},}
\author[18]{Bradley R. Johnson\,\orcidlink{0000-0002-6898-8938},}
\author[1]{Brian Keating\,\orcidlink{0000-0003-3118-5514},}
\author[15]{Andr\'ea Landais\,\orcidlink{0009-0005-8436-1333},}
\author[19]{Toshiya Namikawa\,\orcidlink{0000-0003-3070-9240},}
\author[20]{Federico Nati\,\orcidlink{0000-0002-8307-5088},}
\author[11]{Giuseppe Puglisi\,\orcidlink{0000-0002-0689-4290},}
\author[21]{Christian L. Reichardt\,\orcidlink{0000-0003-2226-9169},}
\author[22]{and Kevin Wolz\,\orcidlink{0000-0003-3155-6151}}

\affiliation[1]{Department of Physics, University of California San Diego,
La Jolla, CA 92093, USA}

\affiliation[2]{Max Planck Institute for Astrophysics,
Karl-Schwarzschild-Str. 1, D-85748 Garching, Germany}

\affiliation[3]{Excellence Cluster ORIGINS,
Boltzmannstr. 2, D-85748 Garching, Germany}

\affiliation[4]{Instituto de Astrof\'isica and Centro de Astro-Ingenier\'ia,
Facultad de F\'isica, Pontificia Universidad Cat\'olica de Chile,
Av. Vicu\~na Mackenna 4860, 7820436 Macul, Santiago, Chile}

\affiliation[5]{Department of Astronomy \& Astrophysics,
University of California San Diego, La Jolla, CA 92093, USA}

\affiliation[6]{Department of Physics, Princeton University -
Jadwin Hall, Princeton University, Princeton, NJ 08544, USA}

\affiliation[7]{The International School for Advanced Studies (SISSA),
via Bonomea 265, I-34136 Trieste, Italy}

\affiliation[8]{The National Institute for Nuclear Physics (INFN),
via Valerio 2, I-34127 Trieste, Italy}

\affiliation[9]{The National Institute for Astrophysics (INAF),
via Tiepolo 11, I-34143 Trieste, Italy}

\affiliation[10]{The Institute for Fundamental Physics of the Universe (IFPU),
Via Beirut 2, I-34151 Trieste, Italy}

\affiliation[11]{Dipartimento di Fisica e Astronomia,
Universit\`a degli Studi di Catania,
via S. Sofia 64, 95123 Catania, Italy}

\affiliation[12]{Department of Physics and Astronomy,
School of Physical, Chemical and Environmental Sciences,
Cardiff University, The Parade, Cardiff, Wales, UK CF243AA}

\affiliation[13]{QUP (WPI), KEK, Tsukuba, Ibaraki 305-0801, Japan}

\affiliation[14]{Department of Physics and Astronomy -
David Rittenhouse Laboratory, Philadelphia, PA 19104}

\affiliation[15]{Universit\'e Paris Cit\'e, CNRS,
Astroparticule et Cosmologie, F-75013 Paris, France}

\affiliation[16]{Department of Physics, University of Texas at Austin,
Austin, TX 78712, USA}

\affiliation[17]{Weinberg Institute for Theoretical Physics,
Texas Center for Cosmology and Astroparticle Physics,
Austin, TX 78712, USA}

\affiliation[18]{Department of Astronomy, University of Virginia,
Charlottesville, VA 22904, USA}

\affiliation[19]{Kavli IPMU (WPI), UTIAS, The University of Tokyo,
Kashiwa, 277-8583, Japan}

\affiliation[20]{Department of Physics, University of Milano-Bicocca,
Piazza della Scienza 3, 20126 Milano, Italy}

\affiliation[21]{School of Physics, The University of Melbourne,
Parkville, VIC 3010, Australia}

\affiliation[22]{Department of Physics, University of Oxford,
Denys Wilkinson Building, Keble Road,
Oxford OX1 3RH, United Kingdom}

\emailAdd{alonappan@ucsd.edu}
\abstract{
Cosmic birefringence, the rotation of the CMB linear polarization plane as photons travel through the Universe, is a powerful test of the existence of parity-violating physics beyond the Standard Model. The measurement of a static isotropic rotation, however, is fundamentally limited by the accuracy of the absolute polarization angle calibration of an instrument. We present a methodology for transferring absolute calibration between instruments of the Simons Observatory (SO), and validate it on map-level simulations of the Large Aperture Telescope (LAT) and Small Aperture Telescopes (SATs) that incorporate lensed CMB polarization, polarized galactic foregrounds, correlated instrumental noise, and polarization-angle miscalibration. We develop two complementary pipelines: a joint LAT+SAT likelihood in which the absolute calibration of either instrument can be transferred to the other through their overlap, and the Minami--Komatsu foreground-based method, which disentangles cosmic birefringence from instrumental miscalibration under assumptions about Galactic dust polarization, without requiring an external instrumental polarization-angle prior. Our aim is to establish how the constraint on the birefringence angle $\beta$ responds to the analysis inputs, mapping its dependence on instrument calibration uncertainty, LAT--SAT overlap multipole range, and analysis strategy. We derive a general variance budget, showing quantitatively that transferring a calibration is only useful for increasing sensitivity when the reference instrument's uncertainty combined with the uncertainty in the transfer using the overlap region is still lower than the calibration uncertainty of the receiving instrument. Transferred calibration does not improve the absolute-angle floor for isotropic $\beta$, but provides an angle reference across the receiving instrument's multipole range. The relative-angle uncertainty is $\sim 0.1^\circ$ for $\ell_{\max}^{\rm SAT}=300$, improving to $\sim 0.04^\circ$ at $\ell_{\max}^{\rm SAT}=600$, subject to SAT beam and transfer-function control. The ultimate uncertainty in $\beta$ depends entirely on the assumptions made about the calibration uncertainties. Finally, for a fiducial $\beta=0.3^\circ$, we show how the SO LAT's $\ell$-range can help distinguish between scale-independent dark-energy-like rotation and scale-dependent axion dark-matter interpretations through the angular dependence of $C_\ell^{EB}$.
}

\begin{document}
\maketitle
\flushbottom
\section{Introduction}
The cosmic microwave background (CMB) provides a powerful observational window into the early universe, offering precise insights into fundamental physics. 
One intriguing phenomenon that can be imprinted in the CMB polarization is \emph{cosmic birefringence} (CB), a frequency-independent rotation of the linear polarization angle of CMB photons as they traverse cosmological distances. 
This effect would indicate the existence of parity-violating interactions and new physics beyond the Standard Model~\citep{Carroll:1989vb,Carroll:1998zi,Carroll:1991zs,Harari:1992ea,Ni:2004jj,Ni:2007ar,Kostelecky:2007zz}.

CB manifests as a rotation of the polarization plane across the sky, which could depend on the observed direction \cite{Lonappan:2025hwz} or be uniform (isotropic) across the entire sky.
 Such an isotropic rotation could arise from a slowly time-evolving and nearly homogeneous pseudoscalar field coupled to photons through a Chern-Simons interaction~\citep{Carroll:1998zi,Ni:2004jj}. 
This type of field is predicted in various theoretical frameworks, including those involving dark energy~\citep{Carroll:1998zi,Fujita:2020aqt,Panda:2010uq,Fujita:2020ecn,Choi:2021aze,Obata:2021nql,Gasparotto:2022uqo,Galaverni:2023zhv}, dark matter~\citep{Finelli:2008jv,Liu:2016dcg,Fedderke:2019ajk}, early dark energy~\citep{Murai:2022zur,Eskilt:2023nxm,Fujita:2020ecn}, topological defects~\citep{Takahashi:2020tqv,Kitajima:2022jzz,Jain:2022jrp,Gonzalez:2022mcx}, and quantum gravity~\cite{Myers:2003fd,Balaji:2003sw,Arvanitaki:2009fg}. 
In many of these models, the field that sources the rotation is that of an axion-like particle (ALP). 
In this case, the birefringence signal offers an indirect probe of the ALP properties, including mass, coupling strength, and cosmological evolution. Searches for a polarization angle oscillating with observation time rely on relative temporal stability; they are distinct from the scale-dependent $EB$ signatures considered here for fields evolving during CMB polarization generation.

The rotation introduced by isotropic CB mixes the parity-even $E$ or $T$ and parity-odd $B$ modes of the CMB polarization~\cite{Kamionkowski:1996ks, Seljak:1996gy}, leading to non-zero $TB$ and $EB$ correlations~\cite{Lue:1998mq,Feng:2004mq}, even if the primordial signal lacks them. 
In particular, among the parity-odd two-point CMB statistics, the $EB$ power spectrum
is a more sensitive probe of birefringence than the $TB$ spectrum, owing to its lower
cosmic variance~\cite{Planck:2016soo}. 
Delensing the $B$-mode field further improves the $EB$-based birefringence sensitivity by reducing the lensing-induced variance~\cite{IdicherianLonappan:2025trj}.
Moreover, if the pseudoscalar field evolves during key epochs such as recombination or reionization, it can subtly alter the shape and amplitude of the $EB$ spectrum~\cite{Galaverni:2023zhv}, increasing the usefulness of this probe in constraining fundamental physics. 
Reconstructing this angular dependence thus offers a tomographic view of the dynamics of the underlying field~\citep{Sigl:2018fba,Nakatsuka:2022epj, Liu:2006uh, Lee:2013mqa, Gubitosi:2014cua, Eskilt:2023nxm, Fujita:2020ecn, Finelli:2008jv}, while also helping to distinguish it from instrumental effects~\citep{QUaD:2008ado, Komatsu:2011a, BICEP1:2013rur, Planck:2016soo, Sherwin:2021vgb, Monelli:2022pru}.

Hints of a non-zero isotropic birefringence angle, $\beta$, have emerged from several CMB experiments. Analyses of data from \textit{Planck}~\citep{Sullivan:2025btc,Minami:2020odp,Diego-Palazuelos:2022dsq,Eskilt:2022wav,Planck:2016soo,Remazeilles:2025wzd}, the Atacama Cosmology Telescope (ACT)~\citep{ACT:2025fju,Diego-Palazuelos_2025}, and joint \textit{Planck}-\textit{WMAP}~\citep{Cosmoglobe:2023pgf,Eskilt:2022cff} and \textit{Planck}-ACT~\citep{Eskilt:2026imm} studies, have reported consistent signals of around $\beta\approx0.2^\circ$ to $0.3^\circ$ at a statistical significance ranging from ${\sim}2\sigma$ in individual analyses to ${\sim}4.8\sigma$ in the combined \textit{ACT} and \textit{Planck} data~\citep{Eskilt:2026imm}. Earlier efforts from QUaD~\citep{QUaD:2008ado}, \textit{WMAP}~\citep{2013ApJS20819H}, POLARBEAR~\citep{POLARBEAR:2019kzz}, the South Pole Telescope~\citep{SPT:2020cxx,PhysRevD.111.083534}, and SPIDER~\cite{SPIDER:2021ncy,Yin:2025fmj} were inconclusive due to systematic uncertainties. Further studies have tried to quantify possible systematic uncertainties affecting these measurements, such as instrumental polarization angles~\citep{Diego-Palazuelos:2022cnh,BICEPKeck:2024cmk}, or the potential intrinsic non-zero $EB$ correlation from Galactic thermal dust~\citep{Clark_2021,Diego-Palazuelos:2022dsq,Cukierman_2023,Hervias_2025}. These hints motivate more sensitive tests with ongoing ground-based experiments such as SO~\citep{SimonsObservatory:2018koc,so_lat_25,SimonsObservatory:2025avm} and AliCPT~\cite{Zhong:2024tgw,Dou:2024wdy,Zhai:2025hqt}, the planned \textit{LiteBIRD} mission~\citep{lb,LiteBIRD:2025mvy}, and the proposed CMB-HD concept~\citep{CMB-HD:2022bsz}, alongside tighter control of systematic uncertainties.

SO is located at 5200\,m of altitude in the Atacama desert in northern Chile, targeting fundamental questions in cosmology. 
It consists of Small Aperture Telescopes (SATs, 0.42\,m aperture) optimized for degree-scale polarization, and a Large Aperture Telescope (LAT, 6\,m aperture), which will target a variety of science cases. 
In this work, we develop and validate the methodology needed to extract the CB signal from SO LAT and SAT data, using map-level simulations of both instruments.

These measurements are ultimately limited by the accuracy of the absolute polarization angle calibration, since a global miscalibration is indistinguishable from an isotropic birefringence rotation~\citep{Kaufman2016Parity}.
Polarization-angle errors also mix $E$ and $B$ modes and can contaminate the inferred cosmological polarization signals~\citep{ODea2007Systematics}. Achieving this calibration to the accuracy required for CB is a central experimental challenge. Ground-based experiments have pursued several complementary strategies: artificial polarized sources such as sparse wire grids and drone- or tower-mounted emitters~\citep{Nati:2017lnn,Coppi_2025,Murata_2023,wiregrid,Dunner:2020}, detailed optical modeling of the instrument-induced polarization rotation~\citep{Koopman:2016abc}, astrophysical calibrators, most notably the Crab nebula (Tau A), the brightest polarized source in the microwave sky~\citep{Aumont:2018epb,Ritacco:2024}, and cross-experiment correlation techniques~\citep{Rigouzzo:2026gzc}. Space-based artificial calibrators have also been proposed, notably CalSat, a CubeSat concept designed to provide a common polarized reference source for ground-based and sub-orbital CMB experiments~\citep{Johnson2015CalSat,Kaufman2016Parity}.
Each of these has practical limits. Wire-grid and artificial-source calibrations determine the instrument's internal polarization response, how a known input angle at the source maps to a detector-frame readout, but the orientation of the source itself relative to the sky, and hence of the instrument boresight, must still be surveyed or otherwise independently established; a wire grid alone cannot verify that this geometric link is correct. The absolute angle of Tau A is currently known only to ${\sim}0.3^\circ$~\citep{Aumont:2018epb}. An alternative is to self-calibrate the polarization angle by requiring the CMB $TB$ and $EB$ spectra to vanish. However, this procedure also removes an isotropic CB signal, while residual Galactic foregrounds with intrinsic non-zero $TB$ or $EB$ correlations can bias the inferred calibration angle~\citep{Abitbol2016Foreground,Kaufman2016Parity}.

These considerations and the difficulty of calibrating any one instrument motivate the strategy explored in this work, which combines calibration measurements for each instrument through the CMB maps the instruments each make of the same sky, as a consistency test and to the benefit of the less-well-calibrated instrument. 

This paper is structured as follows. In \textcolour{blue}{Sec.~\ref{sec:theory}}, we outline the theoretical framework of isotropic CB and its connection to new physics. \textcolour{blue}{Sec.~\ref{sec:simulation}} describes our simulation pipeline, detailing the modeling of astrophysical and cosmological signals and of instrumental effects. In \textcolour{blue}{Sec.~\ref{sec:methods}}, we develop two complementary analysis methodologies and characterize their sensitivity to CB. Finally, in \textcolour{blue}{Sec.~\ref{sec:discussion}}, we discuss the broader implications of our findings and future directions for improving CB constraints.

\section{Theory}\label{sec:theory}

Cosmic birefringence arises from the coupling of a pseudoscalar field $\phi$ to electromagnetism via a Chern-Simons term~\cite{Carroll:1989vb,Carroll:1998zi}. This coupling modifies the photon dispersion relation so that the two circular polarization states propagate with slightly different phase velocities; treating the photon propagation in the geometric-optics (WKB) limit, valid since the photon wavelength is far smaller than the scale over which the field varies, one finds that the birefringent phase difference between the two states accumulates into a net rotation of the linear polarization plane. For a field with a quadratic potential evolving on a $\Lambda$CDM background, the derivation of the photon dispersion relation and the WKB propagation of polarization are standard~\cite{Komatsu:2022nvu}; the key result is that the net rotation of the linear polarization angle accumulated between conformal time $\eta$ and today is
\begin{equation}
\beta(\eta) = \frac{g\,\phi_{\rm in}}{2}\bigl[f(\eta_0)-f(\eta)\bigr],
\label{eq:beta_f_def}
\end{equation}
where $g$ is the coupling constant, $\phi_{\rm in}$ is the initial field value, and $f(\eta)\equiv\phi(\eta)/\phi_{\rm in}$ is the dimensionless evolution function satisfying
\begin{equation}
f'' + 2\frac{a'}{a}\,f' + a^2 m_\phi^2 f = 0,
\label{eq:eom_phi}
\end{equation}
with initial conditions $f(\eta_{\rm in})=1$ and $f'(\eta_{\rm in})=0$ set deep in the radiation era when $H(\eta_{\rm in})\gg m_\phi$. The mass $m_\phi$ governs when the field begins to roll: for $m_\phi\lesssim 10^{-33}$\,eV the field is effectively frozen throughout the history of CMB polarization generation, while for $m_\phi\gtrsim 10^{-28}$\,eV it begins to oscillate before or during recombination~\cite{Murai:2022zur,Nakatsuka:2022epj}. These two regimes produce qualitatively distinct signatures in the CMB $EB$ power spectrum, as described in \textcolour{blue}{Secs.~\ref{subsec:dm_cb}} and~\ref{subsec:const_beta} and illustrated in Fig.~\ref{fig:eb_theory}.

\subsection{Birefringence from a time-dependent field}
\label{subsec:time_dep_cb}

We describe how a time-dependent $\beta(\eta)$ enters the Boltzmann equation for CMB polarization, following Ref.~\cite{Nakatsuka:2022epj}. Solving the line-of-sight integral in the presence of a time-varying rotation rate $\beta'(\eta)=g\phi'(\eta)/2$, the formal solution for the polarization multipole moments at the present time is
\begin{equation}
{}_{\pm2}\Delta_{P,\ell}(\eta_0,q)
= -\frac{3}{4}\sqrt{\frac{(\ell+2)!}{(\ell-2)!}}
\int_0^{\eta_0} d\eta\,
\tau'(\eta)\,e^{-\tau(\eta)}\,
\Pi(\eta,q)\,
\frac{j_\ell[q(\eta_0-\eta)]}{[q(\eta_0-\eta)]^2}\,
e^{\pm 2i\beta(\eta)},
\label{eq:formal_solution}
\end{equation}
where $\tau'(\eta)$ is the differential optical depth, $\Pi(\eta,q)$ is the polarization source term, and $j_\ell$ is the spherical Bessel function of order $\ell$; full definitions follow the conventions of Ref.~\cite{Nakatsuka:2022epj}. The $E$- and $B$-mode transfer functions are obtained from ${}_{\pm2}\Delta_{P,\ell}$, and the $EB$ power spectrum is given by
\begin{equation}
C_\ell^{EB} = 4\pi \int d(\ln q)\,
P_s(q)\,\Delta_\ell^E(q)\,\Delta_\ell^B(q),
\label{eq:cls}
\end{equation}
where $P_s(q)$ is the primordial scalar curvature power spectrum. The factor $e^{\pm 2i\beta(\eta)}$ in Eq.~\eqref{eq:formal_solution} is where the field dynamics enters. CMB polarization is sourced primarily at recombination ($z\simeq 1090$) and reionization ($z\simeq 7$); if $\beta(\eta)$ varies significantly across the widths of these visibility peaks, photons last scattered at different times accumulate different rotation phases, and the resulting $EB$ spectrum develops a characteristic multipole dependence.

\subsubsection{Dark matter regime}
\label{subsec:dm_cb}
\begin{figure}[t]
    \centering
    \includegraphics[width=0.5\linewidth]{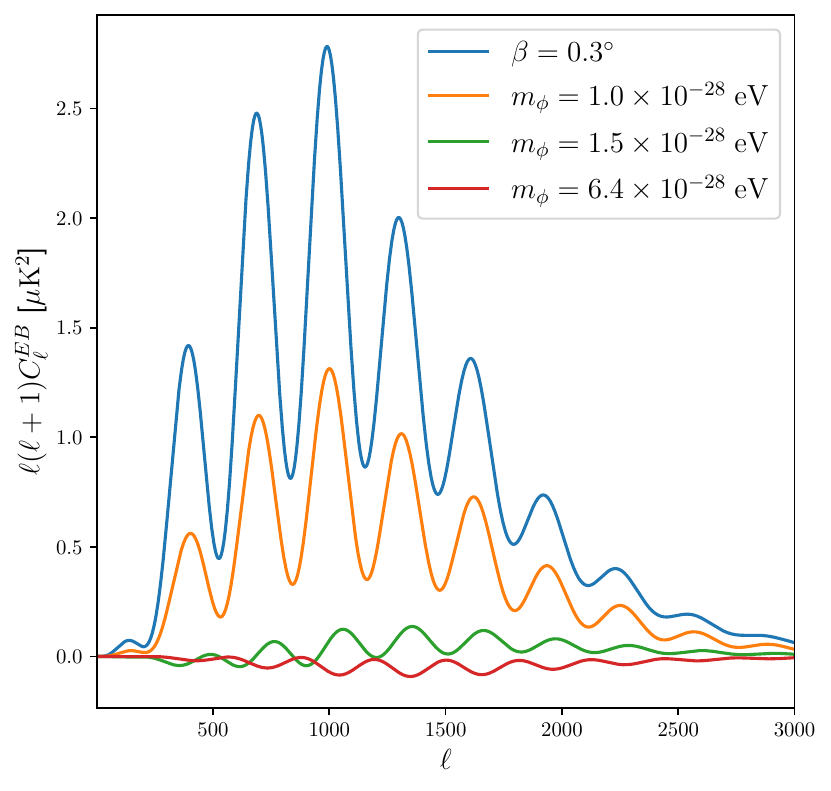}
    \caption{Theoretical $EB$ power spectra for isotropic CB generated by dark energy and dark matter fields. The blue curve shows the spectrum for a dark energy scenario with a rotation of $\beta = 0.3^{\circ}$. The orange, green, and red curves correspond to dark matter fields with masses $m_\phi = 10^{-28}~\mathrm{eV}$, $1.5 \times 10^{-28}~\mathrm{eV}$, and $6.4 \times 10^{-28}~\mathrm{eV}$, respectively, all computed with $\beta(\eta_{\rm in}) = 0.3^{\circ}$}.
    \label{fig:eb_theory}
\end{figure}

If the pseudoscalar field is heavy enough to have begun oscillating by recombination, it behaves as an oscillating dark-matter-like field. For such fields with $m_\phi\gtrsim10^{-28}$\,eV, the field oscillates during recombination, leading to partial cancellations in the net rotation and producing an $EB$ spectrum with oscillatory structure at high multipoles~\cite{Murai:2022zur,Nakatsuka:2022epj}. At the transitional mass range $10^{-32}\,\mathrm{eV}\lesssim m_\phi\lesssim 10^{-31}$\,eV, the field starts evolving after recombination but before reionization, modifying the relative amplitude of the low-$\ell$ reionization bump in $C_\ell^{EB}$~\cite{Sherwin:2021vgb,Nakatsuka:2022epj}. This transitional regime is illustrated separately in Ref.~\cite{Nakatsuka:2022epj}; in this work we instead focus on the oscillatory high-multipole regime, with the theoretical $EB$ spectra for three selected values shown as the orange, green, and red curves in Fig.~\ref{fig:eb_theory}. Masses between $10^{-31}$ and $10^{-28}\,\mathrm{eV}$ interpolate between the regimes discussed here; we do not include them in the fixed-template comparison.

\subsubsection{Dark energy limit}
\label{subsec:const_beta}

If the field is light enough to remain frozen by Hubble friction until the late universe, its nearly constant energy density acts as a dark-energy-like component, and it begins to roll only once $H(\eta)$ drops below $m_\phi$. For such very light fields with $m_\phi\lesssim10^{-33}$\,eV, $f(\eta)$ varies only slowly over both visibility peaks, and the factor $e^{\pm2i\beta(\eta)}$ can be treated as approximately constant under the line-of-sight integral. The resulting scale-independent rotation is shown as the blue curve in Fig.~\ref{fig:eb_theory}. In this limit, the effect of CB reduces to a uniform rotation of the polarization basis by a single effective angle $\beta=g\phi_{\rm in}\Delta f/2$, giving
\begin{align}
C_\ell^{EE,\,\mathrm{obs}} &= \cos^2(2\beta)\,C_\ell^{EE}
            + \sin^2(2\beta)\,C_\ell^{BB},
\label{eq:cee_rot}\\[4pt]
C_\ell^{BB,\,\mathrm{obs}} &= \cos^2(2\beta)\,C_\ell^{BB}
            + \sin^2(2\beta)\,C_\ell^{EE},
\label{eq:cbb_rot}\\[4pt]
C_\ell^{EB,\,\mathrm{obs}} &= \frac{1}{2}\,\sin(4\beta)\,
            \bigl(C_\ell^{EE}-C_\ell^{BB}\bigr),
\label{eq:std_eb}
\end{align}
where the spectra on the right-hand side correspond to the standard $\Lambda$CDM spectra in the absence of birefringence. Since for CMB $C_\ell^{BB}\ll C_\ell^{EE}$, the dominant signature is $C_\ell^{EB,\mathrm{obs}}\approx2\beta C_\ell^{EE}$, corresponding to a scale-independent rotation with amplitude $\beta$. Throughout this work, the map-level simulations of Sec.~\ref{sec:simulation} adopt
this constant-rotation scenario. The dark-matter spectra are not propagated
through the simulation pipeline, but enter solely as fixed
templates in the model-comparison test of Sec.~\ref{sec:de_dm}.

\section{Simulation pipeline}\label{sec:simulation}
In this section, we detail the simulation pipeline used in our analysis, including its components and varying complexity levels. We generate 100 realizations at a \texttt{HEALPix}\,\footnote{\url{http://healpix.sf.net/}}~\cite{2005ApJ...622..759G,Zonca2019} resolution of \texttt{NSIDE} = 2048.\footnote{This resolution is used for the LAT analysis up to $\ell_{\max}=3000$. Although the SAT maps could
in principle be downgraded to a coarser resolution, we retain a uniform \texttt{NSIDE} =
2048 across all maps for computational simplicity. The \texttt{HEALPix} pixelization is adopted for convenience, not in anticipation of the CAR projection of the SO data products.}
The simulation includes signal, noise, and instrumental systematics components as described below. The calibration scenarios explored are summarized in Table~\ref{tab:calibration_cases}.
Other instrumental effects, such as beam asymmetry and ellipticity, gain mismatch, residual temperature-to-polarization leakage, and tube- or focal-plane-dependent angle offsets beyond the single global miscalibration, are not modeled. Map-level analysis can include time-domain effects and transfer functions, but we omit them from the present simulations for computational simplicity. The $E$-to-$B$ leakage induced by the cut sky is accounted for in the power-spectrum estimation with \texttt{NaMaster} (Sec.~\ref{sec:methods}). The results should therefore be read as the sensitivity attainable once these effects are controlled, with the calibration uncertainty taken as the representative dominant systematic.

\subsection{Signal: CMB and foregrounds}\label{sec:signal}

\paragraph{CMB:} We generate lensed CMB polarization maps with \texttt{lenspyx}~\cite{Reinecke:2023gtp}, using unlensed CMB and lensing-potential spectra computed by \texttt{CAMB}~\cite{Lewis:1999bs} for a flat $\Lambda$CDM cosmology with $H_0=67.36\,\mathrm{km\,s^{-1}\,Mpc^{-1}}$, $\Omega_b h^2=0.02237$, $\Omega_c h^2=0.12$, $\Omega_\nu h^2=6.4514\times10^{-4}$, and $\Omega_K=0$. We incorporate cosmic birefringence by rotating the \emph{unlensed} Q and U polarization fields of the CMB by a constant angle $\beta$ prior to lensing. We consider two simulation suites: a signal case with $\beta=0.3^\circ$ and a null case with $\beta=0.0^\circ$ (see Sec.~\ref{subsec:const_beta} for the full scope of birefringence scenarios considered in this work, including the dark-matter templates used separately in Sec.~\ref{sec:de_dm}). The value $\beta=0.3^\circ$ is chosen to be representative of current hints~\cite{Minami:2020odp,Diego-Palazuelos:2022dsq} and is used to test the unbiasedness of the estimators; the sensitivities quoted in this work are derived from the null suite.

\paragraph{Foregrounds:} Galactic foregrounds are simulated using the \texttt{PySM}\,\footnote{\url{https://github.com/galsci/pysm}} package~\cite{Thorne_2017, Zonca_2021, Panexp_2025}. As our baseline we adopt the \texttt{d1s1} combination: the \texttt{d1} thermal-dust template, a single-component modified blackbody with spatially varying spectral index $\beta_d$ and temperature $T_d$ derived from the \textit{Planck} \texttt{Commander} analysis, together with the \texttt{s1} synchrotron template, a power-law model with spatially varying spectral index from the same analysis. To assess robustness to foreground complexity, we additionally consider the \texttt{d10s5} combination, where \texttt{d10} is a GNILC-based dust model~\cite{2016A&A...596A.109P} with small-scale fluctuations used to extend the template to higher multipoles, and \texttt{s5} is a synchrotron model derived from \textit{WMAP} 9-year data at 23\,GHz~\cite{2013ApJS..208...20B} with a spatially varying spectral index consistent with S-PASS observations~\cite{Krachmalnicoff:2018imw}. For the LAT Minami--Komatsu (MK) foreground-based analysis, we further use the alternative \texttt{d9s4} template set when testing template mismatch; \texttt{d9} is its simplified dust component (\texttt{d10} amplitudes with fixed $\beta_d=1.48$ and $T_d=19.6$\,K). The \texttt{s4} component is a simplified version of \texttt{s5} with a fixed synchrotron spectral index of $-3.1$ across the sky.
We note that the \texttt{PySM} \texttt{d1}, \texttt{d9}, and \texttt{d10} models
are not constructed to reproduce the microphysical origin of the dust $EB$ signal, and the residual $EB$ correlation they carry is not physically motivated. Neither \texttt{d1s1} nor \texttt{d10s5} captures filamentary morphology, magnetic misalignment, line-of-sight decorrelation, or spatially varying polarization angles and emission properties~\cite{Clark_2021,Cukierman_2023,Pelgrims:2021gqi,Ritacco:2022qzo,Vacher:2022mvr,Vacher:2022xdw,Hervias_2025}; testing against more sophisticated models~\cite{Vansyngel:2016fbn,Martinez-Solaeche:2017mgz,Vacher:2024adb} is left to future work.
These limitations are especially relevant to the MK analysis, which separates cosmic birefringence from instrumental miscalibration under assumptions about Galactic dust $EB$ polarization.

\paragraph{Beams and passbands:} The LAT and SAT beams are modeled as symmetric
Gaussians following the nominal optical design reported in \cite{so_lat_25} and
\cite{SimonsObservatory:2025avm}, respectively. For the LAT, bandpass integration is performed using the nominal SO passbands. For the SAT, delta bandpasses are adopted for consistency with previous SAT forecasts~\cite{SimonsObservatory:2025avm}.

\subsection{Noise}\label{sec:noise}
Gaussian noise realizations are generated using updated SO noise models that include contributions from white instrumental noise and correlated $1/f$ atmospheric noise, as well as tube-to-tube noise correlations for the LAT. For the SAT, we adopt the noise model of the expanded six-telescope SAT array described in~\cite{SimonsObservatory:2025avm}, which reflects the increased detector count in the 93 and 145\,GHz channels and the survey duration extended to 2035. For the LAT, we use the noise model of the enhanced LAT configuration described in~\cite{so_lat_25}, which incorporates the upgraded 30,000-detector focal plane and the corresponding sensitivity improvements.
We consider the two sensitivity assumptions forecast by SO, denoted \emph{baseline} and \emph{goal}, with the goal scenario corresponding to the more optimistic instrument performance forecasts.
The $1/f$ noise component is fixed to the pessimistic scenario in all cases. Unless stated otherwise, we use the Baseline noise configuration.
The full observation time is divided into two independent observation splits, and all
power spectra entering the analysis are computed as cross-spectra between splits rather
than auto-spectra, in order to avoid the noise bias that would otherwise contaminate the
auto-spectrum estimator. Two splits, rather than the four expected for the LAT, suffice for noise-bias avoidance but give a coarser handle on time-domain systematics. In the noise-dominated limit, using two rather than four splits increases the cross-spectrum variance by a factor of $3/2$; the effect on the full likelihood also depends on signal and correlated noise.
The noise realizations are isotropic. On the smaller LAT--SAT overlap footprint, spatial variations in both noise and Galactic foregrounds, and their interplay, could affect the relative-angle estimate (Sec.~\ref{sec:sat4lat_deltaalpha}); their impact is not quantified in this study.

\subsection{Polarization angle calibration uncertainty}
\label{sec:pol_angle}
Uncertainty in the absolute polarization-angle calibration constitutes one of the dominant instrumental systematics in isotropic birefringence analyses.
In CMB experiments, the absolute angle is calibrated using astrophysical sources and artificial polarized sources.
For SO, the SAT and LAT offer different routes to absolute-angle calibration, each with practical advantages and limitations.
We denote the true band-level polarization-angle offset of instrument $T\in\{\mathrm{LAT},\mathrm{SAT}\}$ by $\alpha_T$, its calibration estimate by $\bar\alpha_T$, and the uncertainty associated with that estimate by $\sigma(\alpha_T)$.

The SAT benefits from dedicated instrumental techniques, including a continuously rotating half-wave plate that helps control instrumental polarization and low-frequency noise that could otherwise complicate polarization-angle measurements, and the ability to deploy artificial calibrators in its far field, which lies at a distance of only a few hundred meters. We stress that the half-wave plate does not by itself set the absolute orientation of the polarization frame on the sky, which still requires an external reference, and that the encoder offset and the plate orientation are themselves terms in the absolute-angle budget. The LAT, whose far field lies tens of kilometers away, is harder to calibrate with artificial sources, but its arcminute resolution gives access to compact polarized astrophysical calibrators and to the high-multipole $EB$ information that drives the statistical constraint.

The transfer techniques developed in Sec.~\ref{sec:sat4lat} are symmetric under LAT$\leftrightarrow$SAT: they apply generically to the comparison of two independent calibrations, and the better-determined angle, whichever instrument it belongs to, can be transferred to the other.
We therefore explore the calibration scenarios of Table~\ref{tab:calibration_cases}, which cover the three possible orderings of the two calibrations. In C1--C3 the SAT is better calibrated than the LAT, and its calibration is transferred to the LAT; C4 assumes comparable calibrations, $\sigma(\alpha_{\rm LAT})=\sigma(\alpha_{\rm SAT})$; and C5 assumes a LAT better calibrated than the SAT. These values span a range chosen to expose the behaviour of the estimators; they should be interpreted as an exploration of possible calibration uncertainties, not as projections of SO performance.

\begin{table}[tb]
\centering
\renewcommand{\arraystretch}{1.2}
\begin{tabular}{lcc}
\hline\hline
Scenario & $\sigma(\alpha_{\rm LAT})$ [deg] & $\sigma(\alpha_{\rm SAT})$ [deg] \\
\hline
C1 & 0.20 & 0.10 \\
C2 & 0.10 & 0.05 \\
C3 & 0.05 & 0.01 \\
C4 & 0.10 & 0.10 \\
C5 & 0.05 & 0.10 \\
\hline\hline
\end{tabular}
\caption{Polarization-angle calibration scenarios adopted in this work. The table lists the assumed Gaussian calibration uncertainties, $\sigma(\alpha)$, on the LAT and SAT polarization angles. Scenarios C2 and C3 assume a better-calibrated SAT, C4 comparable calibrations, and C5 a better-calibrated LAT. Scenario C1 is used only in the exploratory study of Table~\ref{tab:mle_lmax600}. The corresponding choice between transferring the calibration through the LAT--SAT overlap and using each instrument's own prior directly is discussed in Sec.~\ref{sec:sat4lat_regimes}.}
\label{tab:calibration_cases}
\end{table}

We define the SAT absolute polarization angle to be $0^\circ$, so that only the SAT calibration \emph{uncertainty}, $\sigma(\alpha_{\rm SAT})$, enters the analysis as an external prior. The LAT reference angle, then, is allowed to be arbitrary. Its assumed central value $\alpha_{\rm LAT}$ and its uncertainty $\sigma(\alpha_{\rm LAT})$ both enter the simulations; whether $\sigma(\alpha_{\rm LAT})$ also enters the likelihood as a prior depends on the estimator, as described in Sec.~\ref{sec:sat4lat_likelihood}. For each simulation realization and frequency band, we draw the angle offset from a Gaussian centered on the fiducial value in Table~\ref{tab:parameters}, with width $\sigma(\alpha_T)$.

\begin{figure}[t]
    \centering
    \includegraphics[width=0.9\linewidth]{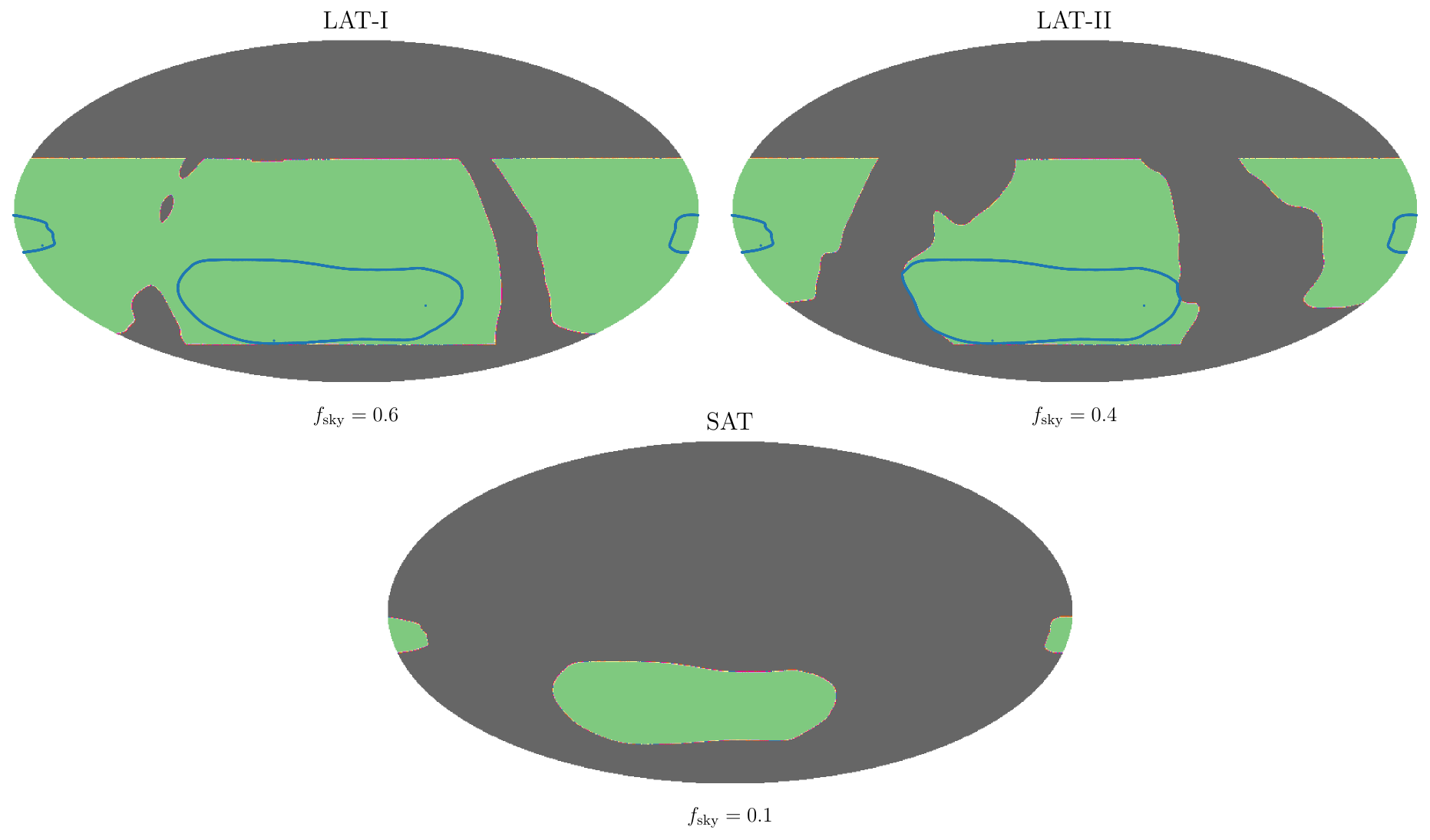}
    \caption{Survey masks derived from the SO mission-scale simulations. \textit{Top left to right:} LAT-I and LAT-II masks; \textit{bottom:} SAT mask. The LAT-I mask corresponds to the LAT footprint with a $10\%$ Galactic cut (baseline), and LAT-II to the same footprint with a $40\%$ Galactic cut (foreground robustness assessment). The $10\%$ ($40\%$) Galactic cut uses a Planck Galactic mask that removes the brightest Galactic region covering $10\%$ ($40\%$) of the full sky, before intersection with the LAT footprint. The SAT mask covers the nominal SAT survey region. Green denotes the retained survey area, gray the masked area, and blue the LAT--SAT overlap contour. The thin transition at the mask boundary reflects the $2^\circ$ apodization. The masks are shown in a Mollweide projection in equatorial coordinates. For the location of the Galactic plane in these coordinates, see Fig.~3 of Ref.~\cite{SimonsObservatory:2018koc}. The LAT$\cap$SAT overlap region has effective sky fraction $f_{\rm sky}\simeq0.1$ and sets the uncertainty on the relative LAT--SAT polarization angle}.
\label{fig:masks}
    \end{figure}

\subsection{Analysis mask}
We adopt three survey masks derived from the SO mission-scale simulations, as shown in Fig.~\ref{fig:masks}. The SAT mask corresponds to the nominal SAT survey footprint and has a sky fraction of $f_{\mathrm{sky}}^{\mathrm{SAT}} = 0.1$~\cite{SimonsObservatory:2025avm}. The two LAT masks are derived from the nominal LAT survey footprint, which has a baseline sky fraction of $f_{\mathrm{sky}}^{\mathrm{LAT}} = 0.61$, and are defined as follows:
\begin{itemize}
    \item \textbf{LAT-I}: the LAT footprint with an additional $10\%$ Galactic cut, retaining the larger fraction of the LAT sky;
    \item \textbf{LAT-II}: the LAT footprint with an additional $40\%$ Galactic cut, providing a more aggressive suppression of polarized Galactic emission.
\end{itemize}
In the LAT cases, regions contaminated by strong CO emission are excluded from the
analysis. For simplicity, individual point sources are not masked; point-source holes would change the mask-induced $E$--$B$ coupling and could affect the estimator variance~\cite{ACT:2025fju}.
The masks are apodized with a $2^\circ$ cosine apodization window following the C2 convention of \texttt{NaMaster}\footnote{\url{https://github.com/LSSTDESC/NaMaster}}~\cite{Alonso:2018jzx}.

The LAT-I mask is adopted as the baseline for both the joint LAT+SAT likelihood analysis of Sec.~\ref{sec:sat4lat} and the Minami--Komatsu (MK) analysis of Sec.~\ref{sec:MLE}. In the joint analysis, the LAT and SAT spectra are computed over the LAT-I and SAT footprints respectively. The LAT$\times$SAT cross-spectra are computed over their common overlap region. In the MK analysis, the full LAT-I mask is preferred as it preserves access to regions of bright Galactic foreground emission that help break the degeneracy between cosmological birefringence and instrumental miscalibration. The LAT-II mask is used as a foreground robustness assessment for the joint LAT+SAT analysis, where the more aggressive Galactic cut quantifies the residual impact of polarized foregrounds on the recovered $\beta$.

\section{Methodology}\label{sec:methods}

In this section, we describe two independent analysis pipelines developed to probe isotropic CB using simulated SO data. The first is the joint LAT+SAT likelihood (Pipeline 1; Sec.~\ref{sec:sat4lat}), which combines both instruments to constrain their polarization angles and the CB angle. The second is the MK estimator (Pipeline 2; Sec.~\ref{sec:MLE})~\cite{2019PTEP.2019h3E02M, 2020PTEP.2020f3E01M, 2020PTEP.2020j3E02M, Diego-Palazuelos:2022cnh}, which uses polarized Galactic foregrounds to separate birefringence from instrumental miscalibration.
The two approaches are compared in Sec.~\ref{sec:method_comparison}.

We deliberately use a simple power-spectrum implementation to study how each method responds to its inputs. It is not a prescription for the pipeline that will be applied to SO data. The likelihood approximations it rests on are stated where they enter (Secs.~\ref{sec:sat4lat_likelihood} and~\ref{sec:MLE}), and the relative trends reported below are more robust to them than the absolute uncertainties.

All power spectra are computed with \texttt{NaMaster} using a uniform bin width of $\Delta\ell = 10$; the pseudo-$C_\ell$ estimator accounts for the mode coupling, including the $E$-to-$B$ leakage, induced by the survey masks. The baseline configuration uses $\ell_{\rm min}^{\rm LAT}=100$ and
$\ell_{\rm max}^{\rm LAT}=3000$, which for $\ell_{\rm min}$ follows the cosmic
birefringence forecast parameters in Ref~\cite{so_lat_25}, and for $\ell_{\rm max}$ adopts a
less conservative cut of 3000 rather than the $\ell_{\rm max}=2000$ used in that
reference, while remaining well within the $\ell=5000$ polarization reach assumed for
other LAT science cases in Ref~\cite{SimonsObservatory:2018koc}. We adopt $\ell_{\max}^{\rm SAT}=300$ as a conservative baseline analysis cut, following the SAT degree-scale polarization setup. It is not a physical limit of the calibration-transfer method; the usable high-$\ell$ range depends on the SAT beam, noise, and transfer-function control, which we explore by varying this cut.

\subsection{Joint LAT+SAT likelihood}
\label{sec:sat4lat}

The joint likelihood compares LAT and SAT measurements of CMB polarization on the same sky and angular scales to determine their relative angle, allowing an absolute calibration from either instrument to be transferred to the other. The transfer is most useful when one instrument has a tighter absolute-angle calibration and the common CMB modes determine the relative angle accurately.
We first develop the analytical framework before applying it to SO simulations in \textcolour{blue}{Sec.~\ref{sec:sat4lat_results}}.

\subsubsection{Polarization rotation and the observed $E$/$B$ spectra}
\label{sec:sat4lat_rotation}

A uniform rotation of the linear polarization by an angle $\theta$ acts on the spin-$\pm2$ combinations of the Stokes parameters as
\begin{equation}
[Q+iU]^{\rm obs}(\hat{n}) \;=\; e^{2i\theta}\,[Q+iU](\hat{n}),
\label{eq:QU_rotation}
\end{equation}
This rotation mixes the $E$- and $B$-mode harmonic coefficients of the CMB signal according to
\begin{align}
E_{\ell m}^{\rm obs} &= \phantom{-}\cos(2\theta)\,E_{\ell m} \;-\; \sin(2\theta)\,B_{\ell m},
\label{eq:E_rot}\\
B_{\ell m}^{\rm obs} &= \phantom{-}\sin(2\theta)\,E_{\ell m} \;+\; \cos(2\theta)\,B_{\ell m}.
\label{eq:B_rot}
\end{align}
We define a positive rotation by the phase $e^{2i\theta}$ in Eq.~\eqref{eq:QU_rotation}; with this convention, positive instrumental miscalibration and birefringence add in $\theta_i=\alpha_i+\beta$, and $\Delta\alpha=\alpha_L-\alpha_S$. The simulated $Q/U$ maps follow the HEALPix/COSMO polarization convention.
For a given map $i$, the total rotation angle receives two physically distinct contributions: a map-specific instrumental miscalibration $\alpha_i$ that needs to be corrected through calibration, assumed constant across the sky within each frequency band, which has uncertainty $\sigma(\alpha_i)$, and the isotropic birefringence angle $\beta$, common to all maps of the same sky. Combining these gives a map-specific rotation angle $\theta_i$:
\begin{equation}
\theta_i \;\equiv\; \alpha_i + \beta .
\label{eq:theta_def}
\end{equation}

Assuming the unrotated CMB is parity-conserving ($C_\ell^{EB}=0$) and
substituting into the cross-spectra between maps $i$ and $j$, where $C_\ell^{EE}$ and
$C_\ell^{BB}$ denote the theoretical lensed CMB spectra, one obtains
\begin{align}
C_\ell^{E_iE_j,\,{\rm obs}}
&= C_\ell^{EE}\cos(2\alpha_i+2\beta)\cos(2\alpha_j+2\beta)
+ C_\ell^{BB}\sin(2\alpha_i+2\beta)\sin(2\alpha_j+2\beta),
\label{eq:cee_obs}\\[2pt]
C_\ell^{B_iB_j,\,{\rm obs}}
&= C_\ell^{EE}\sin(2\alpha_i+2\beta)\sin(2\alpha_j+2\beta)
+ C_\ell^{BB}\cos(2\alpha_i+2\beta)\cos(2\alpha_j+2\beta),
\label{eq:cbb_obs}\\[2pt]
C_\ell^{E_iB_j,\,{\rm obs}}
&= C_\ell^{EE}\cos(2\alpha_i+2\beta)\sin(2\alpha_j+2\beta)
- C_\ell^{BB}\sin(2\alpha_i+2\beta)\cos(2\alpha_j+2\beta).
\label{eq:ceb_obs}
\end{align}
Two structural properties of these expressions are central to what follows. First, all three observed individual instrument spectra depend on $\beta$ only through the symmetric combination $\theta_i+\theta_j = \alpha_i+\alpha_j+2\beta$. Second, there are combinations of cross spectra that depend only on $\theta_i-\theta_j = \alpha_i-\alpha_j$, independent of $\beta$.

In their present form, Eqs.~(\ref{eq:cee_obs})-(\ref{eq:ceb_obs}) consider only the CMB signal. Therefore, their validity is limited to scenarios where Galactic foreground emission makes a subdominant contribution. The results in \textcolour{blue}{Sec.~\ref{sec:sat4lat_results}} show that, for common models such as \texttt{d1s1}, the SO SAT and LAT observations at 93 and 145\,GHz fulfill such a condition within the angular scale cuts and masking strategy adopted in our analysis. The simulation tests include polarized Galactic foregrounds, so they capture foreground-induced effects to the extent represented by the adopted PySM models. As Galactic emission might prove more complex than the models adopted in our simulation study, we further assess the impact of foreground contamination and explore methodology extensions to marginalize over it in \textcolour{blue}{Sec.~\ref{sec:sat4lat_results}}~\cite{Jost:2022oab}.

\subsubsection{Uncertainty on total rotation from the $EB$ likelihood}
\label{sec:sat4lat_thetaL}

Setting $i=j=L$ in Eq.~\eqref{eq:ceb_obs} gives the LAT auto-$EB$ spectrum,
\begin{equation}
C_\ell^{E_LB_L,\,{\rm obs}}
\;=\;
\tfrac{1}{2}\sin\bigl(4[\alpha_L+\beta]\bigr)
\bigl(C_\ell^{EE}-C_\ell^{BB}\bigr),
\label{eq:ceb_LL}
\end{equation}
which for the small rotations of interest reduces to
\begin{equation}
C_\ell^{E_LB_L,\,{\rm obs}} \;\simeq\; 2(\alpha_L+\beta)\bigl(C_\ell^{EE}-C_\ell^{BB}\bigr).
\label{eq:ceb_LL_small}
\end{equation}
The LAT auto-spectra therefore constrain only the single combination
\begin{equation}
\theta_L \;=\; \alpha_L + \beta,
\label{eq:thetaL_def}
\end{equation}
with no internal handle on the relative contributions of $\alpha_L$ and $\beta$. We denote the LAT-internal uncertainty on this combination by $\sigma(\theta_L)$; the analogous SAT-internal quantity $\theta_S = \alpha_S + \beta$ has uncertainty $\sigma(\theta_S)$. We use $L$ and $S$ as shorthand for LAT and SAT in the equations; thus $\alpha_L=\alpha_{\rm LAT}$, $\alpha_S=\alpha_{\rm SAT}$, and $\widehat\beta_L=\widehat\beta_{\rm LAT}$, with analogous notation for SAT.

\subsubsection{LAT--SAT relative angle from the overlap cross-spectra}
\label{sec:sat4lat_deltaalpha}

Forming the symmetric and antisymmetric combinations of the LAT$\times$SAT cross-spectra ($L\neq S$) from Eqs.~\eqref{eq:cee_obs}--\eqref{eq:ceb_obs} yields
\begin{align}
C_\ell^{E_LE_S,\,{\rm obs}} + C_\ell^{B_LB_S,\,{\rm obs}}
&= \bigl(C_\ell^{EE}+C_\ell^{BB}\bigr)\cos\bigl(2[\alpha_L-\alpha_S]\bigr),
\label{eq:sym_combo}\\[2pt]
C_\ell^{E_SB_L,\,{\rm obs}} - C_\ell^{E_LB_S,\,{\rm obs}}
&= \bigl(C_\ell^{EE}+C_\ell^{BB}\bigr)\sin\bigl(2[\alpha_L-\alpha_S]\bigr).
\label{eq:antisym_combo}
\end{align}
Two features of these expressions are essential. First, the CMB dependence enters only through the rotation-invariant sum $C_\ell^{EE}+C_\ell^{BB}$. Second, $\beta$ cancels identically, leaving only the relative LAT--SAT miscalibration angle
\begin{equation}
\Delta\alpha \;\equiv\; \alpha_L - \alpha_S.
\label{eq:deltaalpha_def}
\end{equation}
The triplet $\{\theta_L,\theta_S,\Delta\alpha\}$ exhausts the information available
from CMB polarization alone in the LAT--SAT system: $\theta_L$ and $\theta_S$ are the
only combinations extractable from the LAT and SAT auto-spectra respectively
(Sec.~\ref{sec:sat4lat_thetaL}), and $\Delta\alpha$ is the only combination extractable
from their cross-spectra (this section): every bilinear
combination of rotated $E$/$B$ harmonics reduces to a function of $\theta_i\pm\theta_j$
(Sec.~\ref{sec:sat4lat_rotation}), so no further independent rotation observable exists. These three observables
obey the consistency relation $\theta_L-\theta_S=\Delta\alpha$ and therefore span only a
two-dimensional subspace of the three-dimensional parameter space $(\alpha_L,\alpha_S,
\beta)$: $\beta$ remains formally degenerate with a common shift of both miscalibration
angles, and an external absolute-angle reference is required to break this degeneracy.
Combining the two telescopes does not lift it either:
what the combination adds is statistical lever arm, the LAT multipole range sharpening
$\theta_L$ and the overlap sharpening $\Delta\alpha$, so that once an absolute prior is imposed on at least one instrument
(Sec.~\ref{sec:sat4lat_likelihood}) the constraint on $\beta$ may then improve as compared to  either
telescope alone.

The cross-spectrum combinations above represent one of several possible estimators of $\Delta\alpha$; a map-level alternative is given in Appendix~\ref{sec:app_map_calib} and a $TB$-based one in Appendix~\ref{sec:ext_TB}; these provide cross-checks of the recovered relative angle, with different uncertainties discussed in the appendices.
The cross-spectrum estimator is adopted as the baseline throughout this work.

\subsubsection{Joint LAT+SAT likelihood and $\beta$ estimators}
\label{sec:sat4lat_likelihood}

The parameters $\boldsymbol{\Theta}\equiv(\theta_L,\theta_S,\Delta\alpha)$ are determined jointly from a Gaussian bandpower likelihood combining three observational blocks plus external absolute-angle calibration priors on the two instruments,
\begin{equation}
-2\ln\mathcal{L}(\boldsymbol{\Theta},\alpha_S,\alpha_L)
\;=\;
\chi^2_{\rm EB}(\theta_L,\theta_S)
+\chi^2_{\rm sym}(\Delta\alpha)
+\chi^2_{\rm anti}(\Delta\alpha)
+\chi^2_{\rm prior}(\alpha_S,\alpha_L).
\label{eq:logL_total}
\end{equation}
We fit $\Delta\alpha$ independently of $\theta_L$ and $\theta_S$, rather than impose $\theta_L-\theta_S=\Delta\alpha$, so their agreement remains an internal consistency test. Imposing the relation would make the transferred LAT-side $\beta$ estimator algebraically equivalent to the direct SAT-side estimator.

The first block uses the telescope-internal $EB$ cross-spectra summed over all same-telescope map pairs $(i,j)$ with $i\neq j$ and multipole bins $\ell\in\mathcal{L}_T$ for each telescope $T\in\{L,S\}$. Here $i$ and $j$ label independent data splits of the same frequency channel, so that the sum contains no noise-biased auto-spectra; cross-frequency $EB$ spectra, such as $93\times145$, are not included, and the per-frequency columns of Table~\ref{tab:mle_results} are therefore separately fitted single-channel constraints, whose errors can be correlated through their shared CMB sky,
 \begin{equation}
\chi^2_{\rm EB}(\theta_L,\theta_S)
 \;=\;
 \sum_{T\in\{L,S\}}\sum_{i\neq j}\sum_{\ell\in\mathcal{L}_T}
\frac{\bigl[\widehat{C}_\ell^{E_iB_j}-C_\ell^{E_iB_j}(\theta_T,\theta_T)\bigr]^2}
     {\sigma^2\bigl[\widehat{C}_\ell^{E_iB_j}\bigr]},
\label{eq:chi2_EB}
\end{equation}
where the model spectrum is given by Eq.~\eqref{eq:ceb_obs} evaluated at the lensed $\Lambda$CDM input spectra. The second and third blocks use the LAT$\times$SAT cross-spectra over the common multipole range $\mathcal{L}_{LS}$, restricted to same-frequency pairs, implementing the symmetric and antisymmetric combinations of Eqs.~\eqref{eq:sym_combo}--\eqref{eq:antisym_combo}. We restrict the LAT--SAT comparison to same-frequency pairs so that common-sky foreground signals cancel in the relative-angle combination; cross-frequency pairs would require an explicit model of foreground frequency scaling and its covariance. With two splits per instrument, the relative-angle likelihood includes all four LAT$\times$SAT split pairs at each frequency; the map-level check in Appendix~\ref{sec:app_map_calib} uses same-index pairs only. The likelihood terms are
\begin{align}
\chi^2_{\rm sym}(\Delta\alpha) &= \sum_{\ell\in\mathcal{L}_{LS}}
\frac{\bigl[\widehat{X}_\ell-X_\ell^{\rm th}\cos(2\Delta\alpha)\bigr]^2}{\sigma^2[\widehat{X}_\ell]},
\label{eq:chi2_sym}\\[2pt]
\chi^2_{\rm anti}(\Delta\alpha) &= \sum_{\ell\in\mathcal{L}_{LS}}
\frac{\bigl[\widehat{Y}_\ell-X_\ell^{\rm th}\sin(2\Delta\alpha)\bigr]^2}{\sigma^2[\widehat{Y}_\ell]},
\label{eq:chi2_anti}
\end{align}
with measured bandpowers
\begin{equation}
\widehat{X}_\ell \equiv \widehat{C}_\ell^{E_LE_S}+\widehat{C}_\ell^{B_LB_S},
\qquad
\widehat{Y}_\ell \equiv \widehat{C}_\ell^{E_SB_L}-\widehat{C}_\ell^{E_LB_S},
\qquad
X_\ell^{\rm th} \equiv C_\ell^{EE}+C_\ell^{BB}.
\label{eq:Xhat_def}
\end{equation}
All bandpower variances are estimated from the Monte Carlo simulation set of \textcolour{blue}{Sec.~\ref{sec:simulation}}. We retain only the variance of each bandpower, omitting covariance between multipole bins and between frequency channels; the likelihood also neglects covariance between its observational blocks. Because the frequency bands share CMB sample variance, frequency-combined uncertainties obtained without their cross-covariance are approximate and should not be interpreted as gains from independent measurements. The 100-simulation estimate is treated as exact. The fourth block contains the Gaussian polarization-angle calibration priors of the two instruments,
\begin{equation}
\chi^2_{\rm prior}(\alpha_S,\alpha_L) \;=\; \frac{(\alpha_S-\bar{\alpha}_S)^2}{\sigma^2(\alpha_S)}
\;+\;\frac{(\alpha_L-\bar{\alpha}_L)^2}{\sigma^2(\alpha_L)},
\label{eq:chi2_prior}
\end{equation}
with central values $\bar{\alpha}_T$ and widths $\sigma(\alpha_T)$ provided by independent calibration campaigns external to the CMB analysis (see \textcolour{blue}{Sec.~\ref{sec:pol_angle}}). An instrument without a usable absolute calibration corresponds to the limit $\sigma(\alpha_T)\to\infty$, in which its term drops out of Eq.~\eqref{eq:chi2_prior}.

The joint likelihood determines the three independent combinations
\begin{equation}
\theta_L=\alpha_L+\beta,\qquad
\theta_S=\alpha_S+\beta,\qquad
\Delta\alpha=\alpha_L-\alpha_S,
\label{eq:three_obs}
\end{equation}
which form a rank-two system in the three unknowns $(\alpha_L,\alpha_S,\beta)$; the missing direction is fixed by the calibration priors. The absolute angle of each instrument can then be reconstructed along two routes: \emph{directly}, from its own prior, or by \emph{transfer}, from the prior of the other instrument through the relative angle, $\alpha_L=\alpha_S+\Delta\alpha$ or $\alpha_S=\alpha_L-\Delta\alpha$. Subtracting the reconstructed angle from the corresponding total rotation gives, for each telescope, a direct and a transferred estimator of the birefringence angle,
\begin{align}
\widehat{\beta}^{\,\rm dir}_{\rm LAT} &= \theta_L-\alpha_L,
&\widehat{\beta}^{\,\rm tr}_{\rm LAT} &= \theta_L-\alpha_S-\Delta\alpha,
\label{eq:beta_estimator}\\
\widehat{\beta}^{\,\rm dir}_{\rm SAT} &= \theta_S-\alpha_S,
&\widehat{\beta}^{\,\rm tr}_{\rm SAT} &= \theta_S-\alpha_L+\Delta\alpha.
\label{eq:beta_estimator_sat}
\end{align}
The transferred estimators use the LAT--SAT overlap, while the direct estimators do not. Both are limits of the weighted estimator introduced in Sec.~\ref{sec:sat4lat_regimes}, which also gives the combination of the LAT- and SAT-side estimates into a single estimator $\widehat{\beta}$.

\subsubsection{Variance budget and limiting regimes}
\label{sec:sat4lat_regimes}

We first write the variance budget in its most general form and then specialise it to the three possible orderings of the two calibrations. Let $T\in\{L,S\}$ denote the telescope for which we intend to improve the absolute angle calibration using telescope $\bar T$. The absolute angle of $T$ can be estimated as a weighted combination of the direct and transferred routes of Sec.~\ref{sec:sat4lat_likelihood},
\begin{equation}
\widehat{\alpha}_T(w_T) \;=\; w_T\,\alpha_T \;+\; (1-w_T)\,\bigl(\alpha_{\bar T}+s_T\,\Delta\alpha\bigr),
\qquad s_L=+1,\;\; s_S=-1,
\label{eq:alpha_weighted}
\end{equation}
with $0\le w_T\le1$, giving the estimator $\widehat{\beta}_T(w_T)=\theta_T-\widehat{\alpha}_T(w_T)$. The direct and transferred estimators of Eqs.~\eqref{eq:beta_estimator}--\eqref{eq:beta_estimator_sat} correspond to $w_T=1$ and $w_T=0$, respectively. The ingredients $\theta_T$, $\Delta\alpha$, $\alpha_L$ and $\alpha_S$ come from the telescope-internal $EB$ block, the overlap blocks and the two external priors, and are independent at lowest order. Linear error propagation then gives
\begin{equation}
\sigma^2\bigl(\widehat{\beta}_T\bigr)
\;=\;
\sigma^2(\theta_T)
\;+\; w_T^2\,\sigma^2(\alpha_T)
\;+\; (1-w_T)^2\bigl[\sigma^2(\alpha_{\bar T})+\sigma^2(\Delta\alpha)\bigr].
\label{eq:beta_variance_general}
\end{equation}
The optimal weight $w_T^\star$, which minimises Eq.~\eqref{eq:beta_variance_general}, is
$w_T^\star=\bigl[\sigma^2(\alpha_{\bar T})+\sigma^2(\Delta\alpha)\bigr]/\bigl[\sigma^2(\alpha_T)+\sigma^2(\alpha_{\bar T})+\sigma^2(\Delta\alpha)\bigr]$. At this weight, the minimum variance is
\begin{equation}
\sigma^2_{\rm opt}\bigl(\widehat{\beta}_T\bigr)
= \sigma^2(\theta_T)
+ \left[\frac{1}{\sigma^2(\alpha_T)}
       +\frac{1}{\sigma^2(\alpha_{\bar T})+\sigma^2(\Delta\alpha)}\right]^{-1}.
\label{eq:beta_variance_twosided}
\end{equation}
The two anchors combine harmonically: the tighter one dominates, and the gain from adding the second is bounded, reaching at most $\sqrt{2}$ when the two routes are equally precise. The transferred route is preferable to the direct one whenever
\begin{equation}
\sigma^2(\alpha_{\bar T})+\sigma^2(\Delta\alpha) \;<\; \sigma^2(\alpha_T),
\label{eq:transfer_condition}
\end{equation}
that is, transferring a calibration pays only when the reference instrument calibration uncertainty combined with the relative-angle uncertainty $\sigma(\Delta\alpha)$ delivered by the overlap is still less than the uncertainty in the instrument to which we are transferring the calibration.

For the SO LAT, the mode count at multipoles where the $EB$ likelihood is most sensitive yields $\sigma(\theta_L)\ll\sigma(\alpha_T),\sigma(\Delta\alpha)$ for all cases considered here, so $\sigma(\theta_L)$ is a negligible contribution in every case below. We stress that $\sigma(\theta_L)\simeq0.003^\circ$ is not comparable to the $0.036^\circ$--$0.041^\circ$ of the MK likelihood or the $0.04^\circ$ of Ref.~\cite{so_lat_25}: $\theta_L$ is a single rotation parameter fitted at fixed cosmology and foreground model, whereas those are errors on $\beta$ marginalized over the miscalibration angles and foreground amplitudes that break the degeneracy.

Once the calibration is transferred, the LAT- and SAT-side estimates for $\beta$ are finally combined as $\widehat{\beta}=\lambda\,\widehat{\beta}_L+(1-\lambda)\,\widehat{\beta}_S$. With $\sigma^2_T\equiv\sigma^2(\widehat{\beta}_T)$ and $C\equiv{\rm Cov}(\widehat{\beta}_L,\widehat{\beta}_S)$, the optimal weight gives
\begin{equation}
\sigma^2\bigl(\widehat{\beta}\bigr) \;=\; \frac{\sigma^2_L\,\sigma^2_S - C^2}{\sigma^2_L+\sigma^2_S-2C},
\label{eq:beta_comb}
\end{equation}
with
\begin{equation}
C \;=\; w_L(1-w_S)\,\sigma^2(\alpha_L) + (1-w_L)\,w_S\,\sigma^2(\alpha_S) - (1-w_L)(1-w_S)\,\sigma^2(\Delta\alpha),
\label{eq:beta_cov}
\end{equation}
The covariance arises because the two sides can share the same prior or relative-angle information. For two direct estimators ($w_L=w_S=1$) the covariance vanishes and Eq.~\eqref{eq:beta_comb} reduces to the inverse-variance combination. When the LAT-side estimate is instead transferred from the SAT ($w_L=0$, $w_S=1$), the two sides share the SAT prior, $C=\sigma^2(\alpha_S)$, and Eq.~\eqref{eq:beta_comb} becomes
$\sigma^2(\widehat{\beta})\approx\sigma^2(\widehat{\beta}_L)=\sigma^2(\alpha_S)+\bigl[1/(\sigma^2(\theta_L)+\sigma^2(\Delta\alpha))+1/\sigma^2(\theta_S)\bigr]^{-1}$: the shared absolute calibration cannot be averaged down by combining the two sides. We now specialise Eqs.~\eqref{eq:beta_variance_general}--\eqref{eq:beta_comb} to the calibration scenarios of Table~\ref{tab:calibration_cases}.

\paragraph{SAT better calibrated, $\sigma(\alpha_S)<\sigma(\alpha_L)$ (C2, C3).}
This is the configuration in which the SAT calibration is transferred to the LAT. For the transferred fits in C2 and C3, no LAT prior is imposed in the likelihood, $\sigma(\alpha_L)\to\infty$ in Eq.~\eqref{eq:chi2_prior}, so that $w_L^\star=0$ and Eq.~\eqref{eq:beta_variance_general} reduces to the transferred LAT-side budget
\begin{equation}
\sigma^2(\widehat{\beta}^{\,\rm tr}_{\rm LAT})
\;=\;
\sigma^2(\theta_L) \;+\; \sigma^2(\alpha_S) \;+\; \sigma^2(\Delta\alpha),
\label{eq:beta_variance}
\end{equation}
so that the SAT absolute calibration sets a strict floor on the achievable birefringence uncertainty,
\begin{equation}
\sigma(\widehat{\beta}^{\,\rm tr}_{\rm LAT}) \;\geq\; \sigma(\alpha_S).
\label{eq:beta_floor}
\end{equation}
The budget then separates into two limiting regimes set by the relative size of $\sigma(\alpha_S)$ and $\sigma(\Delta\alpha)$. When $\sigma(\alpha_S)\gg\sigma(\Delta\alpha)$, the variance is dominated by the absolute calibration term, $\sigma(\widehat{\beta}_{\rm LAT})\simeq\sigma(\alpha_S)$, and the only lever for improvement is tightening the SAT polarization-angle calibration. In the complementary overlap-limited regime, $\sigma(\Delta\alpha)\gg\sigma(\alpha_S)$, the bottleneck shifts to the relative angle determination, and gains come from extending the overlap multipole range, increasing the LAT--SAT overlap sky fraction, or mitigating foreground $EB$ contamination. The transition between the two regimes is set by
\begin{equation}
\sigma(\alpha_S) \;=\; \sigma(\Delta\alpha).
\label{eq:crossover}
\end{equation}
For C2 and C3 the SAT prior spans $\sigma(\alpha_S)\in[0.01^\circ,0.05^\circ]$, while the baseline overlap configuration delivers $\sigma(\Delta\alpha)\simeq0.1^\circ$--$0.15^\circ$ at 93 and 145\,GHz. Both scenarios therefore sit in the overlap-limited regime. Equation~\eqref{eq:transfer_condition} also shows that the transfer is worthwhile only if the LAT's own calibration, were it used, satisfies $\sigma(\alpha_L)\gtrsim\sqrt{\sigma^2(\alpha_S)+\sigma^2(\Delta\alpha)}\simeq0.11^\circ$--$0.15^\circ$ at the baseline cuts.

\paragraph{Comparable calibrations, $\sigma(\alpha_S)=\sigma(\alpha_L)\equiv\sigma_\alpha$ (C4).}
Substituting into Eq.~\eqref{eq:transfer_condition} gives $\sigma_\alpha^2+\sigma^2(\Delta\alpha)<\sigma_\alpha^2$, which is never satisfied: for comparable calibrations the transferred route is always less precise than the direct one, by the relative-angle uncertainty added in quadrature, and transferring the calibration in either direction is suboptimal. The optimal combination of Eq.~\eqref{eq:beta_variance_twosided} gives
\begin{equation}
\sigma^2_{\rm opt}\bigl(\widehat{\beta}_T\bigr)
=\sigma^2(\theta_T)+\sigma_\alpha^2\,\frac{\sigma_\alpha^2+\sigma^2(\Delta\alpha)}{2\sigma_\alpha^2+\sigma^2(\Delta\alpha)},
\label{eq:beta_variance_equal}
\end{equation}
which approaches the $\sqrt{2}$ gain only for $\sigma(\Delta\alpha)\ll\sigma_\alpha$. At the baseline cuts $\sigma(\Delta\alpha)\gtrsim\sigma_\alpha$, so the overlap adds little, and each instrument is best anchored by its own prior, $w_L=w_S=1$, with $\sigma^2(\widehat{\beta}_T)=\sigma^2(\theta_T)+\sigma_\alpha^2$ and uncorrelated LAT- and SAT-side estimates.

\paragraph{LAT better calibrated, $\sigma(\alpha_L)<\sigma(\alpha_S)$ (C5).}
The roles of the two instruments are now exchanged. Equation~\eqref{eq:transfer_condition} with $T=L$ cannot be satisfied, so the LAT-side estimate should use the LAT prior directly, $\sigma^2(\widehat{\beta}^{\,\rm dir}_{\rm LAT})=\sigma^2(\theta_L)+\sigma^2(\alpha_L)$, which combines the LAT statistical power with the tighter of the two calibrations. Since $\Delta\alpha$ is a relative angle and Eqs.~\eqref{eq:sym_combo}--\eqref{eq:antisym_combo} are symmetric under $L\leftrightarrow S$, the transfer can instead be run in the reverse direction, using the LAT as the absolute reference for the SAT through $\widehat{\beta}^{\,\rm tr}_{\rm SAT}$ of Eq.~\eqref{eq:beta_estimator_sat}, with
\begin{equation}
\sigma^2(\widehat{\beta}^{\,\rm tr}_{\rm SAT})
\;=\;
\sigma^2(\theta_S) \;+\; \sigma^2(\alpha_L) \;+\; \sigma^2(\Delta\alpha),
\label{eq:beta_variance_reverse}
\end{equation}
the mirror image of Eq.~\eqref{eq:beta_variance}. By Eq.~\eqref{eq:transfer_condition} this reverse transfer improves on the SAT's own prior only if $\sigma^2(\alpha_L)+\sigma^2(\Delta\alpha)<\sigma^2(\alpha_S)$. For C5 at the baseline cuts, $\sqrt{0.05^2+0.11^2}\simeq0.12^\circ>\sigma(\alpha_S)=0.1^\circ$, so it does not; it becomes beneficial once the overlap is sharpened (Sec.~\ref{sec:sat4lat_results}).

\paragraph{Consistency test.}
The independent LAT and SAT absolute-angle calibrations can be checked against the relative angle measured from their common sky: $\theta_L-\theta_S=\Delta\alpha$. This check is available whether or not both calibration priors are used in the birefringence likelihood; its sensitivity depends on their uncertainties. When both instruments carry an absolute prior, the system is over-determined and the consistency relation becomes an explicit null test~\citep{Lonappan:2026unb}: the SAT-anchored reconstruction of the LAT angle, $\bar\alpha_S+\widehat{\Delta\alpha}$, can be compared with the LAT's own prior central value $\bar\alpha_L$, normalised by the quadrature sum of $\sigma(\alpha_L)$, $\sigma(\alpha_S)$ and $\sigma(\Delta\alpha)$. A miscalibrated instrument then registers as a tension between the two inputs rather than being absorbed into the combined estimate. The sensitivity of the test is set by whichever term dominates that sum; under the baseline cuts $\sigma(\Delta\alpha)$ does, so the test is limited by the relative-angle determination rather than by the priors themselves.

\subsubsection{Application to SO simulations}
\label{sec:sat4lat_results}

We now apply the formalism developed in this section to the SO map-level simulations described in Sec.~\ref{sec:simulation}. We restrict the analysis to the two science frequencies, 93 and 145\,GHz, where the polarization signal-to-noise is largest and the contribution from polarized Galactic foregrounds is minimal. Additional bands could be included by assigning each band its own angle parameter and accounting for its foreground contribution; they are outside the joint analysis presented here. We sample the joint likelihood of Eq.~\ref{eq:logL_total} with the \texttt{emcee} MCMC sampler~\cite{Foreman-Mackey:2012any}. We consider calibration scenarios C2--C5, listed in Table~\ref{tab:calibration_cases} and discussed in Sec.~\ref{sec:sat4lat_regimes}. For each scenario C2--C5, we evaluate transferred and direct configurations. In the transferred configuration, only the SAT prior is imposed and the LAT is calibrated through the overlap. In the direct configuration, both priors of Eq.~\eqref{eq:chi2_prior} are imposed and each instrument is anchored by its own prior. The main parameters entering the fits, with their input values and priors, are summarized in Table~\ref{tab:parameters}.

\begin{table}[t]
\centering
\renewcommand{\arraystretch}{1.2}
\small
\begin{tabular}{llll}
\hline\hline
Parameter & Pipeline & Input value & Prior \\
\hline
$\beta$ & both & $0.3^\circ$ / $0^\circ$ & derived (joint) / flat (MK) \\
$\theta_L,\ \theta_S,\ \Delta\alpha$ & joint & -- & flat \\
$\alpha_{\rm SAT}$ & joint & $0^\circ$ & $\mathcal{N}(0,\sigma^2(\alpha_{\rm SAT}))$ \\
$\alpha_{\rm LAT}$ & joint & $0.2^\circ$ & flat / $\mathcal{N}(\bar\alpha_{\rm LAT},\sigma^2(\alpha_{\rm LAT}))$ \\
$\alpha_i$ & MK & $-0.1^\circ$/$0.2^\circ$/$0.15^\circ$ & flat \\
$A^{\rm fg}_{{\rm LAT},\nu}$ & joint & PySM template & uniform on $[0,2]$ \\
$A_{\rm d}$ & MK & PySM template & free \\
\hline\hline
\end{tabular}
\caption{Main parameters of the joint LAT+SAT and MK analyses, with their input values and priors. The widths $\sigma(\alpha_{\rm SAT})$ and $\sigma(\alpha_{\rm LAT})$ for each scenario are given in Table~\ref{tab:calibration_cases}. The LAT prior is flat in the transferred configuration and Gaussian in the direct one. The MK input angles refer to the LF/MF/HF bands. The foreground amplitudes are free only in the foreground-robustness tests of the joint pipeline and in the MK pipeline.}
\label{tab:parameters}
\end{table}

\paragraph{Baseline parameter recovery.}
Table~\ref{tab:mle_results} summarizes the $1\sigma$ uncertainties on the fitted parameters $\theta_S$, $\theta_L$, $\Delta\alpha$, and on the derived birefringence estimators $\beta_{\rm SAT}$ and $\beta_{\rm LAT}$, at both frequencies and for all calibration scenarios, together with their frequency-combined values. Several features are worth noting.

First, $\sigma(\theta_L)\ll\sigma(\theta_S)$ at both frequencies by an order of magnitude, confirming that the LAT dominates the total-rotation determination, as anticipated in Sec.~\ref{sec:sat4lat_regimes}. The fitted $\theta_T$ are essentially unchanged between the transferred and direct configurations: the two differ only in how the absolute angle is anchored.

Second, the SAT-side estimator $\beta_{\rm SAT}$ is firmly in the SAT-prior-limited regime: $\sigma(\beta_{\rm SAT})$ tracks $\sigma(\alpha_S)$ across all scenarios, and is identical in C4 and C5, which share $\sigma(\alpha_S)=0.1^\circ$.

Third, and most informative, the transferred LAT-side estimator behaves qualitatively differently: $\sigma(\beta_{\rm LAT})$ remains overlap dominated between C2 and C3 despite a factor-of-five tightening of the SAT prior, tracking the corresponding $\sigma(\Delta\alpha)$ from the overlap blocks. This is the overlap-limited regime of Eq.~\eqref{eq:beta_variance}. Decomposing Eq.~\eqref{eq:beta_variance} into fractional contributions, $\sigma^2(\theta_L)$ is under $1\%$ of $\sigma^2(\beta_{\rm LAT})$ in all scenarios, while $\sigma^2(\Delta\alpha)$ accounts for about $83\%$--$90\%$ in C2 and $\gtrsim98\%$ in C3.

\begin{table*}
\centering
\caption{$1\sigma$ uncertainties (in degrees) on the polarization-angle parameters recovered from the joint LAT+SAT likelihood (Eq.~\ref{eq:logL_total}) under the calibration scenarios C2--C5 of Table~\ref{tab:calibration_cases}, for simulations with input cosmic birefringence $\beta=0^\circ$. $\theta_S$ and $\theta_L$ are the fitted total rotation angles; $\Delta\alpha$ is the LAT--SAT relative miscalibration angle from the overlap blocks (Eqs.~\ref{eq:chi2_sym}--\ref{eq:chi2_anti}); $\beta_{\rm SAT}$ and $\beta_{\rm LAT}$ are the derived cosmic birefringence estimates from the SAT-side and LAT-side estimators of Eqs.~\eqref{eq:beta_estimator}--\eqref{eq:beta_estimator_sat}. \emph{Tr.}: the LAT is calibrated by transfer from the SAT, with no LAT prior imposed; \emph{Dir.}: both priors are imposed and each instrument is anchored by its own prior, so that $\Delta\alpha$ does not enter the estimators. The last three columns give the 93+145\,GHz combination of the SAT-side and LAT-side estimators and their inverse-variance combination $\beta$, computed neglecting the covariance of Eq.~\eqref{eq:beta_comb}. All entries use the baseline multipole cuts $\ell_{\rm min}^{\rm LAT}=100$, $\ell_{\rm max}^{\rm SAT}=300$.}
\label{tab:mle_results}
\renewcommand{\arraystretch}{1.3}
\resizebox{\textwidth}{!}{%
\begin{tabular}{@{}
  ll
  cc
  cc
  cc
  cc
  cc
  ccc
@{}}
\hline\hline
&
& \multicolumn{2}{c}{$\sigma(\theta_S)$}
& \multicolumn{2}{c}{$\sigma(\theta_L)$}
& \multicolumn{2}{c}{$\sigma(\Delta\alpha)$}
& \multicolumn{2}{c}{$\sigma(\beta_{\rm SAT})$}
& \multicolumn{2}{c}{$\sigma(\beta_{\rm LAT})$}
& \multicolumn{3}{c}{93+145\,GHz}
\\
\cmidrule(lr){3-4}
\cmidrule(lr){5-6}
\cmidrule(lr){7-8}
\cmidrule(lr){9-10}
\cmidrule(lr){11-12}
\cmidrule(lr){13-15}
Calibration & Anchoring
& 93 & 145
& 93 & 145
& 93 & 145
& 93 & 145
& 93 & 145
& $\sigma(\beta_{\rm SAT})$ & $\sigma(\beta_{\rm LAT})$ & $\sigma(\beta)$
\\
\hline
\multicolumn{15}{@{}l}{\emph{SAT better calibrated}} \\
C2 & Tr. & 0.02 & 0.03 & 0.003 & 0.003 & 0.11 & 0.15 & 0.05 & 0.06 & {0.12} & {0.16} & 0.04 & 0.10 & 0.04 \\
{C2} & {Dir.} & {0.02} & {0.03} & {0.003} & {0.003} & {--} & {--} & {0.05} & {0.06} & {0.10} & {0.10} & {0.04} & {0.07} & {0.04} \\
C3 & Tr. & 0.02 & 0.03 & 0.003 & 0.004 & 0.11 & 0.15 & 0.02 & 0.03 & 0.11 & 0.15 & 0.02 & 0.09 & 0.02 \\
{C3} & {Dir.} & {0.02} & {0.03} & {0.003} & {0.004} & {--} & {--} & {0.02} & {0.03} & {0.05} & {0.05} & {0.02} & {0.04} & {0.02} \\
\hline
\multicolumn{15}{@{}l}{\emph{Comparable calibrations}} \\
C4 & Tr.  & 0.02 & 0.03 & 0.004 & 0.005 & 0.11 & 0.15 & 0.10 & 0.11 & 0.15 & 0.18 & 0.07 & 0.11 & 0.06 \\
C4 & Dir. & 0.02 & 0.03 & 0.004 & 0.005 & --   & --   & 0.10 & 0.11 & 0.10 & 0.10 & 0.07 & 0.07 & 0.05 \\
\hline
\multicolumn{15}{@{}l}{\emph{LAT better calibrated}} \\
C5 & Tr.  & 0.02 & 0.03 & 0.004 & 0.005 & 0.11 & 0.15 & 0.10 & 0.11 & 0.15 & 0.18 & 0.07 & 0.11 & 0.06 \\
C5 & Dir. & 0.02 & 0.03 & 0.004 & 0.005 & --   & --   & 0.10 & 0.11 & 0.05 & 0.05 & 0.07 & 0.04 & 0.03 \\
\hline\hline
\end{tabular}%
}
\end{table*}

\paragraph{Combined estimator.}
The frequency-combined columns of Table~\ref{tab:mle_results} illustrate the two limiting regimes of \textcolour{blue}{Sec.~\ref{sec:sat4lat_regimes}}. In the SAT-better scenarios the SAT-side estimator scales directly with the SAT prior width: $\sigma(\beta_{\rm SAT})$ improves from $\simeq0.04^\circ$ in C2 to $\simeq0.02^\circ$ in C3, and the transferred combined uncertainty from $\sigma(\beta)\simeq0.04^\circ$ to $\simeq0.02^\circ$. The transferred LAT-side estimator remains close to $\sigma(\beta_{\rm LAT})\sim0.09^\circ$--$0.10^\circ$, reflecting the overlap-dominated budget set by $\sigma(\Delta\alpha)$. Its central value still relies on $\alpha_S$ through Eq.~\eqref{eq:beta_estimator}; it therefore serves as a cross-check, distinct from the foreground-model-dependent MK method of \textcolour{blue}{Sec.~\ref{sec:MLE}}. At the baseline cuts, the direct LAT prior yields a smaller LAT-side uncertainty than SAT-to-LAT transfer in C2 and C3, even though the SAT prior is tighter. Transfer still enables a calibration-consistency comparison. Because the transferred LAT-side and the SAT-side estimators share $\alpha_S$, the combined values for these configurations are lower bounds (Eq.~\eqref{eq:beta_comb}).

We note that $\sigma(\beta_{\rm SAT})$ can fall below the nominal single-frequency $\sigma(\alpha_S)$ of Table~\ref{tab:calibration_cases}; this comes from combining the 93 and 145\,GHz channels, which tightens the \emph{statistical} uncertainty by roughly $\sqrt{2}$, not from any LAT information entering the SAT-side estimator. The same holds for the direct LAT-side estimator, whose frequency-combined uncertainty is $\simeq\sigma(\alpha_L)/\sqrt{2}$. This scaling assumes the two band-level uncertainties to be independent, which holds for the noise-dominated statistical component but not for the systematic terms of the absolute-angle budget. Contributions such as the polarization-angle knowledge of the calibration source, the source and telescope geometry, the HWP encoder offset, and the detector-angle convention are common-mode across bands and do not average down. Writing $\sigma^2(\alpha_T)=\sigma^2_{\rm corr}+\sigma^2_{\rm uncorr}$, the two-band combination gives $\sigma^2_{\rm corr}+\sigma^2_{\rm uncorr}/2$, so the effective gain interpolates between $\sqrt{2}$ and unity. The scenarios of Table~\ref{tab:calibration_cases} should therefore be read as specifying the \emph{total} combined angle uncertainty entering each prior, irrespective of its partition.

\paragraph{Comparable and LAT-favoured calibrations.}
The C4 and C5 rows of Table~\ref{tab:mle_results} test the predictions of Eqs.~\eqref{eq:transfer_condition}--\eqref{eq:beta_variance_reverse}. In the transferred configuration the two scenarios give identical results, $\sigma(\beta_{\rm LAT})=0.15^\circ$/$0.18^\circ$ at 93/145\,GHz, since the transferred estimator does not use the LAT prior and C4 and C5 share the same SAT prior. The transfer route therefore delivers $\sqrt{\sigma^2(\alpha_S)+\sigma^2(\Delta\alpha)}\simeq0.15^\circ$--$0.18^\circ$ per band, regardless of how well the LAT itself is calibrated. Anchoring the LAT on its own prior instead brings $\sigma(\beta_{\rm LAT})$ down to $0.10^\circ$ per band in C4 and $0.05^\circ$ in C5, i.e. to $\sigma(\alpha_L)$, as Eq.~\eqref{eq:beta_variance_general} with $w_L=1$ requires. In the frequency combination, $\sigma(\beta_{\rm LAT})$ improves from $0.11^\circ$ to $0.07^\circ$ in C4 and to $0.04^\circ$ in C5, and the combined $\sigma(\beta)$ from $0.06^\circ$ to $0.05^\circ$ and $0.03^\circ$, respectively. When the calibrations are comparable or the LAT is the better-calibrated instrument, transferring the SAT calibration to the LAT is thus suboptimal, and the direct estimator should be used. In C5 the reverse transfer of Eq.~\eqref{eq:beta_variance_reverse} would not improve the SAT side at the baseline cuts, as anticipated in Sec.~\ref{sec:sat4lat_regimes}.

\begin{figure}
\centering
\includegraphics[width=0.8\columnwidth]{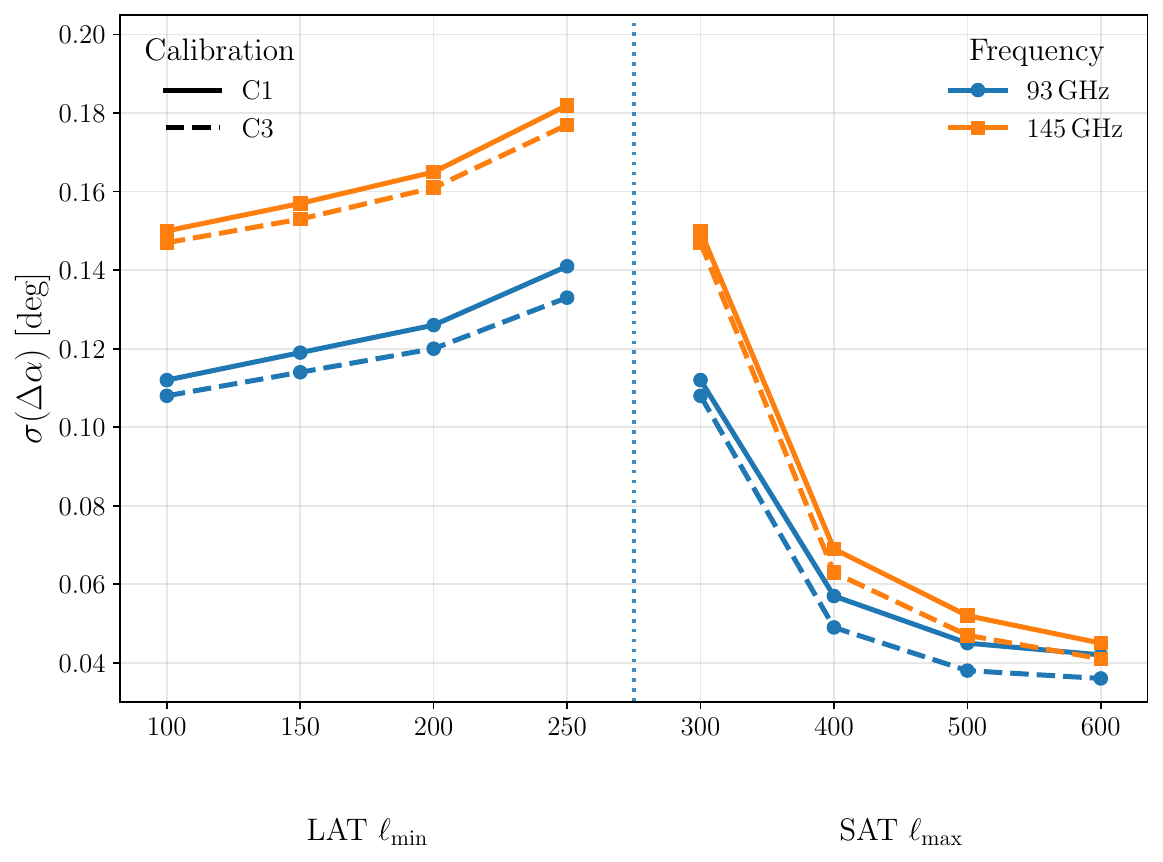}
\caption{Dependence of the LAT--SAT relative-angle uncertainty $\sigma(\Delta\alpha)$ from the joint overlap estimator on the analysis multipole cuts. Left half: scan over $\ell_{\rm min}^{\rm LAT}$ at fixed $\ell_{\rm max}^{\rm SAT}=300$. Right half: scan over $\ell_{\rm max}^{\rm SAT}$ at fixed $\ell_{\rm min}^{\rm LAT}=100$. The vertical dotted line separates the two scans. Solid and dashed lines correspond to calibration scenarios C1 and C3, respectively; colours and markers distinguish 93\,GHz and 145\,GHz. These are null-signal simulations with $\beta=0^\circ$, SO Baseline noise, and \texttt{d1s1} foregrounds. The small C1--C3 differences reflect their separately simulated and estimated bandpower covariances, including scenario-dependent angle realizations; the relative-angle likelihood has no direct dependence on the absolute-angle prior width.}
\label{fig:sigma_alpha_vs_lmin_lmax}
\end{figure}

\paragraph{Dependence on the multipole cuts.}\label{sec:robustness_lmin}
Figure~\ref{fig:sigma_alpha_vs_lmin_lmax} shows how $\sigma(\Delta\alpha)$ responds to variations in the two multipole cuts that define the LAT--SAT overlap. Scanning $\ell_{\rm min}^{\rm LAT}\in\{100,150,200,250\}$ at fixed $\ell_{\rm max}^{\rm SAT}=300$ broadens $\sigma(\Delta\alpha)$ by only $\sim25\%$ across the full range, indicating that the LAT low-$\ell$ modes carry little relative-angle information. This is an important practical advantage: raising $\ell_{\rm min}^{\rm LAT}$ to suppress atmospheric or $1/f$ systematics incurs negligible cost in $\sigma(\Delta\alpha)$. In contrast, scanning $\ell_{\rm max}^{\rm SAT}\in\{300,400,500,600\}$ at fixed $\ell_{\rm min}^{\rm LAT}=100$ produces a factor of $\sim2.7$ improvement in $\sigma(\Delta\alpha)$ at 93\,GHz, identifying the SAT high-$\ell$ reach as the primary lever for reducing the overlap-limited floor on $\sigma(\beta_{\rm LAT})$. In the limiting case where $\ell_{\rm min}^{\rm LAT}>\ell_{\rm max}^{\rm SAT}$, the
LAT and SAT multipole ranges no longer overlap, $\Delta\alpha$ cannot be determined from the cross-spectra, and the best achievable constraint on $\beta$ reduces to $\widehat{\beta}_{\rm SAT}$ alone, limited entirely by the SAT calibration prior $\sigma(\alpha_S)$. We interpret this result as taking advantage of the multipole ranges where the highest signal is: as can be seen in Fig.~\ref{fig:eb_theory}, there is a peak of $EB$ signal under an isotropic rotation around $\ell \sim 400-500$, while at $\ell<250$, there is little signal.

\paragraph{Exploratory extension of the SAT multipole reach.}
To quantify the gain available from the $\ell_{\rm max}^{\rm SAT}$ lever identified above, we re-run the joint inference with $\ell_{\rm max}^{\rm SAT}=600$, keeping all other settings fixed. As summarized in Table~\ref{tab:mle_lmax600}, $\sigma(\Delta\alpha)$ tightens by a factor of $\sim3$ at both frequencies. The impact on $\sigma(\beta_{\rm LAT})$ depends on the calibration scenario in a way that maps directly onto the two regimes of Sec.~\ref{sec:sat4lat_regimes}. For C1, where $\sigma(\alpha_S)\simeq0.1^\circ$ now dominates over the reduced $\sigma(\Delta\alpha)$, the LAT-side estimator crosses into the SAT-prior-limited regime, and $\sigma(\beta_{\rm LAT})$ saturates at $\simeq0.1^\circ$ regardless of further improvements in the overlap. For C3, where $\sigma(\alpha_S)\simeq0.01^\circ$, the system approaches the crossover of Eq.~\eqref{eq:crossover}, yielding $\sigma(\beta_{\rm LAT})\simeq0.038^\circ$/$0.042^\circ$ at 93/145\,GHz.

A sharper overlap also changes the conclusions for C4 and C5. With $\sigma(\Delta\alpha)\simeq0.04^\circ$, Eq.~\eqref{eq:beta_variance_equal} predicts that optimally combining the two routes in C4 would reduce the per-band uncertainty from $\sigma_\alpha=0.1^\circ$ to $\simeq0.07^\circ$, so the overlap becomes a useful second anchor rather than a penalty. In C5, Eq.~\eqref{eq:beta_variance_reverse} gives $\sqrt{\sigma^2(\alpha_L)+\sigma^2(\Delta\alpha)}\simeq0.065^\circ<\sigma(\alpha_S)=0.1^\circ$, so the reverse transfer of the LAT calibration to the SAT would then improve the SAT-side estimate. These C4/C5 optimal-combination and reverse-transfer values are analytic expectations, distinct from the transferred LAT-side pipeline results in Table~\ref{tab:mle_lmax600}. Reaching $\ell_{\rm max}^{\rm SAT}=600$ requires SAT beam and transfer-function control that is not addressed by the map-level simulations used here.

\begin{table*}
\centering
\caption{Exploratory transferred joint LAT+SAT analysis with $\ell_{\rm max}^{\rm SAT}=600$ and $\ell_{\rm min}^{\rm LAT}=100$. $\sigma(\Delta\alpha)$ and $\sigma(\beta_{\rm LAT})$ are $1\sigma$ uncertainties at 93 and 145\,GHz as defined in Table~\ref{tab:mle_results}. The C1--C5 scenarios are evaluated on null-signal simulations with $\beta=0^\circ$, SO Baseline noise, and \texttt{d1s1} foregrounds. The baseline analysis uses $\ell_{\rm max}^{\rm SAT}=300$; these values indicate the gain available from extending the SAT analysis to higher multipoles, subject to additional systematic control not addressed here.}
\label{tab:mle_lmax600}
\renewcommand{\arraystretch}{1.4}
\resizebox{0.5\textwidth}{!}{%
\begin{tabular}{@{}
  l
  cc
  cc
@{}}
\hline\hline
& \multicolumn{2}{c}{$\sigma(\Delta\alpha)$ [deg]}
& \multicolumn{2}{c}{$\sigma(\beta_{\rm LAT})$ [deg]}
\\
\cmidrule(lr){2-3}
\cmidrule(lr){4-5}
Calibration
& 93\,GHz & 145\,GHz
& 93\,GHz & 145\,GHz
\\
\hline
C1 & 0.042 & 0.045 & 0.108 & 0.110\\
{C2} & {0.0375} & {0.0426} & {0.0627} & {0.0659}\\
C3 & 0.036 & 0.041 & 0.038 & 0.042\\
{C4} & {0.0385} & {0.0432} & {0.1073} & {0.1091}\\
{C5} & {0.0379} & {0.0424} & {0.1070} & {0.1087}\\
\hline\hline
\end{tabular}%
}
\end{table*}

\begin{figure*}
\centering
\includegraphics[width=0.6\textwidth]{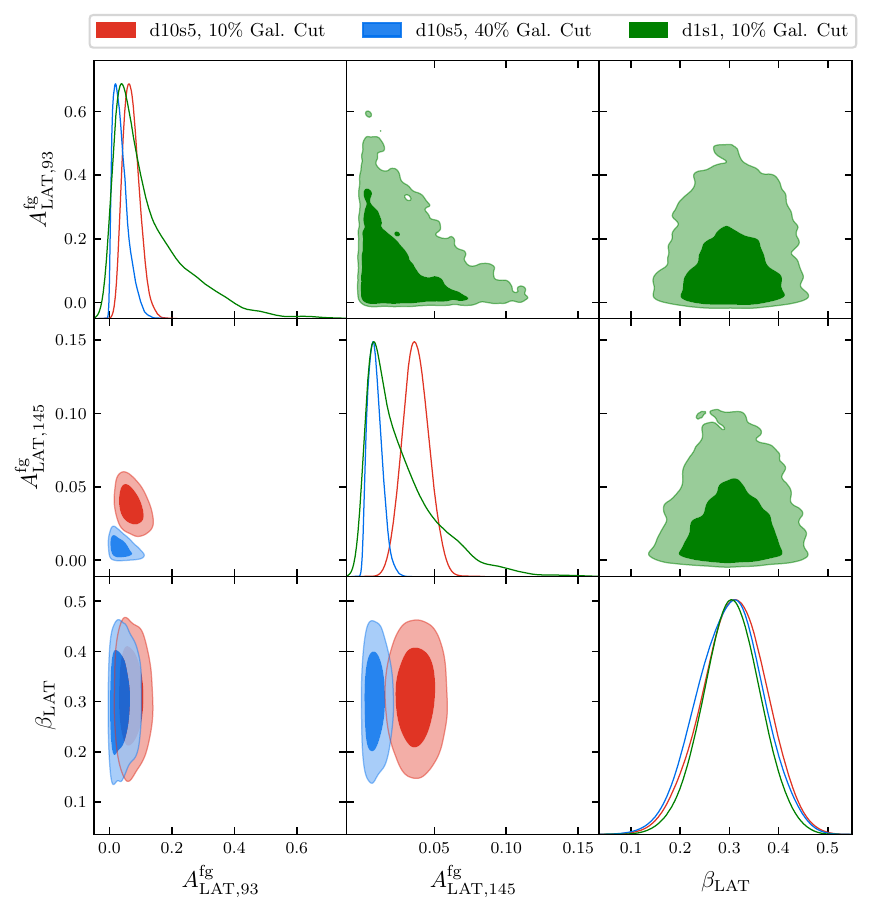}
\caption{Joint posterior distributions of the foreground amplitudes on the $EB$ spectrum $A^{\rm fg}_{{\rm LAT},93}$, $A^{\rm fg}_{{\rm LAT},145}$, and the LAT-side birefringence angle $\beta_{\rm LAT}$ for the joint LAT+SAT estimator in scenario C3 at 93 and 145\,GHz. The configurations use \texttt{d10s5} on the LAT-I ($10\%$ Galactic cut) mask, \texttt{d10s5} on the LAT-II ($40\%$ Galactic cut) mask, and \texttt{d1s1} on the LAT-I mask. All use SO Goal noise. Contours enclose $68\%$ and $95\%$ of the posterior mass; the fiducial value is $\beta=0.3^\circ$.}
\label{fig:fg_impact_triangle}
\end{figure*}

\paragraph{Foreground robustness.}\label{sec:fg_amplitude}
We test the robustness of the LAT-side estimator against polarized Galactic emission by
augmenting the likelihood with per-frequency foreground amplitude parameters
$A^{\rm fg}_{{\rm LAT},\nu}$ ($\nu\in\{93,145\}$\,GHz), scaling a dust-only $EB$
template computed on the full sky from the same \texttt{PySM} foreground model used for the corresponding simulation. At each frequency the model is
\begin{equation}
{C_{\ell,\nu}^{EB,\mathrm{model}}=C_{\ell,\nu}^{EB,\mathrm{CMB}}(\theta_L)+A^{\rm fg}_{{\rm LAT},\nu}C_{\ell,\nu}^{EB,\mathrm{dust,template}}.}
\label{eq:fg_eb_model}
\end{equation}
The template is the unmasked, dust-only PySM spectrum, fitted separately at 93 and 145\,GHz. This term models intrinsic dust $EB$ in the telescope-internal $EB$ spectra. We do not include rotation-induced dust $EB$ from the foreground $EE-BB$ spectra, and the LAT$\times$SAT relative-angle blocks retain their CMB-only model. Figure~\ref{fig:fg_impact_triangle} shows the joint posterior of $A^{\rm fg}_{{\rm LAT},93}$, $A^{\rm fg}_{{\rm LAT},145}$, and $\beta_{\rm LAT}$ for three configurations: the \texttt{d10s5} model on the LAT-I and LAT-II masks, and the baseline \texttt{d1s1} model on LAT-I.

For \texttt{d10s5} on LAT-I, the fitted amplitudes are $A^{\rm fg}_{{\rm LAT},\nu}<0.05$ relative to the full-sky dust $EB$ template. Because the fit uses masked spectra, these coefficients should not be interpreted as the fraction of dust emission remaining on the observed sky. Applying the more aggressive LAT-II mask reduces the foreground contribution further, and the amplitudes become consistent with zero, returning only upper limits. For the baseline \texttt{d1s1} model on LAT-I, the amplitudes are weakly constrained and bounded below by zero, consistent with the weak intrinsic $EB$ signal in this model. The apparent widths of the foreground-amplitude posteriors should be interpreted with the nonnegative prior in mind: when a fit approaches $A^{\rm fg}=0$, truncation shapes the contours and can make an upper-limit posterior appear narrow. We therefore do not interpret narrower contours near zero as evidence for a more precise foreground measurement. The foreground amplitudes have uniform priors on $0\leq A^{\rm fg}\leq2$; the posterior boundary at zero is imposed by this prior.

In the transferred C3 configuration of Fig.~\ref{fig:fg_impact_triangle}, the foreground amplitude enters the LAT total-rotation fit, for which $\sigma(\theta_L)\simeq0.003^\circ$, while $\sigma^2(\Delta\alpha)$ accounts for $\gtrsim98\%$ of the $\beta_{\rm LAT}$ variance. The weak variation of the marginalized $\beta_{\rm LAT}$ posterior with $A^{\rm fg}$ is therefore largely expected from the variance budget. The recovered $\beta_{\rm LAT}$ values agree within the simulated foreground models and mask choices; these tests do not establish foreground robustness beyond those models.

\begin{figure}
\centering
\includegraphics[width=0.4\columnwidth]{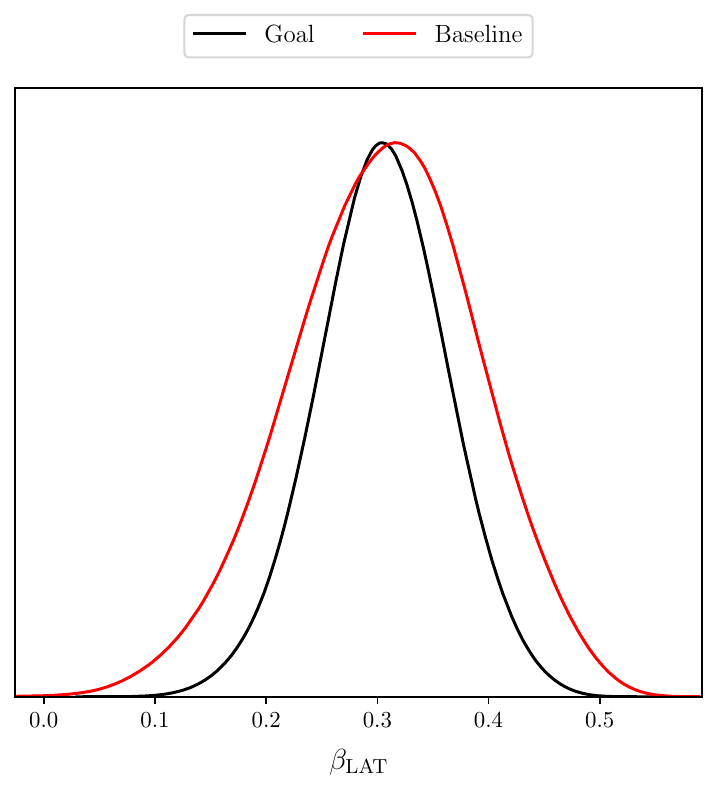}
\caption{Marginalized posterior on $\beta_{\rm LAT}$ from the joint LAT+SAT LAT-side estimator in scenario C3, using 93 and 145\,GHz with \texttt{d1s1} foregrounds. The two curves compare SO Baseline and Goal noise. The fiducial value is $\beta=0.3^\circ$.}
\label{fig:noise_impact_1d}
\end{figure}

\paragraph{Impact of the noise configuration.}
Figure~\ref{fig:noise_impact_1d} compares the marginalized posterior on $\beta_{\rm LAT}$ for the SO Baseline and Goal noise configurations, with a fiducial $\beta=0.3^\circ$. Changing from Baseline to Goal noise reduces $\sigma(\beta_{\rm LAT})$ from $0.09^\circ$ to $0.06^\circ$, driven by a $25$--$30\%$ improvement in $\sigma(\Delta\alpha)$ at both frequencies. This $\sim1.4\times$ tightening is a direct consequence of the overlap-limited scaling $\sigma(\beta_{\rm LAT})\simeq\sigma(\Delta\alpha)$: reduced map noise propagates into the cross-spectrum signal-to-noise that determines $\sigma(\Delta\alpha)$, while the assumed SAT angle-calibration prior is held fixed between the two noise configurations.

\subsection{LAT angle calibration using foregrounds}\label{sec:MLE}

We now discuss the MK estimator, which simultaneously constrains $\beta$ and the per-frequency focal-plane polarization angles $\alpha_i$ across the LAT bands, without an external instrument-angle prior but conditional on an assumed Galactic dust $EB$ model.
The estimator is applied to all six LAT frequency channels, each with two independent noise splits, and uses the LAT-I analysis mask, which preserves the larger sky fraction needed by the foreground-assisted MK estimator.

In this subsection, $i$ and $j$ label frequency bands; the split indices used in the joint likelihood do not appear in these expressions. Following the formulation of \cite{Diego-Palazuelos:2022dsq,Diego-Palazuelos:2022cnh}, the observed $EB$ cross-spectra between a pair of frequency bands ($i,j$) in the presence of CB and miscalibrations can be expressed as

\begin{align}\label{eq: obs eb mk}
C_\ell^{E_iB_j,\,\mathrm{obs}} = &
\phantom{+}\frac{\sin(4\alpha_j)}{\cos(4\alpha_i)+\cos(4\alpha_j)} ~C_\ell^{E_iE_j,\,\mathrm{obs}}
-  \frac{\sin(4\alpha_i)}{\cos(4\alpha_i)+\cos(4\alpha_j)}~C_\ell^{B_iB_j,\,\mathrm{obs}} \nonumber\\
& + \frac{2\cos(2\alpha_i)\cos(2\alpha_j)}{\cos(4\alpha_i)+\cos(4\alpha_j)}~C_\ell^{E_iB_j,\,\mathrm{dust}}
+ \frac{2\sin(2\alpha_i)\sin(2\alpha_j)}{\cos(4\alpha_i)+\cos(4\alpha_j)}~C_\ell^{B_iE_j,\,\mathrm{dust}} \\
&+ \frac{\sin(4\beta)}{2\cos(2\alpha_i+2\alpha_j)} \,b_\ell^ib_\ell^j\omega_\ell^i\omega_\ell^j\left(C_\ell^{EE} - C_\ell^{BB}\right), \nonumber
\end{align}

where $C_\ell^{EE}$ and $C_\ell^{BB}$ are the standard $\Lambda$CDM predictions convolved with the corresponding beam ($b_\ell$), and pixel window functions ($\omega_\ell$) of each band, and $C_\ell^{EB,\,\mathrm{dust}}$ gives the intrinsic $EB$ correlation from polarized Galactic dust emission~\cite{Huffenberger:2019mjx, Clark_2021}. Equation~(\ref{eq: obs eb mk}) highlights the need for modeling polarized dust emission~\cite{Diego-Palazuelos:2022dsq}. Here, we will take the approach of \cite{Diego-Palazuelos:2022cnh,LiteBIRD:2025mvy}, and model the dust $EB$ spectrum from the Galactic emission templates on \texttt{PySM}, therefore substituting $C_\ell^{EB,\,\mathrm{dust}}\rightarrow A_\mathrm{d}C_\ell^{EB,\,\mathrm{dust\,temp.}}$, where $C_\ell^{EB,\,\mathrm{dust\,temp.}}$ is a fixed dust $EB$ template spectrum computed from the \texttt{PySM} maps and $A_\mathrm{d}$ is a free amplitude that rescales it. The dust-template spectra are computed with the same apodized LAT-I mask as the observed spectra. Because this template carries a free amplitude, the non-zero dust $EB$ present in the \texttt{PySM} simulations is absorbed by $A_\mathrm{d}$ rather than biasing the MK estimate. In principle, polarized synchrotron emission may also contribute to Eq.~(\ref{eq: obs eb mk}).
However, no parity-violating emission mechanism for Galactic synchrotron has been proposed to date, and no significant global synchrotron $EB$ correlation has been reported~\cite{Martire:2021gbc,Rubino-Martin:2023fya}; we neglect the synchrotron contribution and assume the intrinsic synchrotron $EB$ signal to be null.
In this section, we initially set $C_\ell^{EB,\mathrm{dust\,temp.}}$ to match the fiducial dust model used in the simulations, but at the end we discuss that the results are not significantly changed by adjusting the template to a different foreground model. The alternative \texttt{d9s4} template set includes synchrotron emission, but the likelihood does not fit a separate intrinsic synchrotron $EB$ term.

A Gaussian likelihood is then constructed,
\begin{equation}\label{eq: mk likelihood}
-2 \ln \mathcal{L} =
\sum_{\ell=\ell_{\min}}^{\ell_{\max}}
\left(\mathbf{A}\,\vec{C}_\ell^{\,\mathrm{obs}} -
\mathbf{B}\,\vec{C}_\ell^{\,\mathrm{CMB}}-\mathbf{F}\vec{C}_\ell^{\,\mathrm{dust}}\right)^{\sf T}
\mathbf{M}_\ell^{-1}
\left(\mathbf{A}\,\vec{C}_\ell^{\,\mathrm{obs}} -
\mathbf{B}\,\vec{C}_\ell^{\,\mathrm{CMB}}-\mathbf{F}\vec{C}_\ell^{\,\mathrm{dust}}\right),
\end{equation}
from the one-dimensional arrays
\begin{align}
  \vec{C}_\ell^{\,\mathrm{obs}} = &~ \big(C_\ell^{E_iE_j},C_\ell^{B_iB_j}, C_\ell^{E_iB_j}\big)^{\sf T}, \\
  \vec{C}_\ell^{\,\mathrm{CMB}} = &~ \big(b_\ell^ib_\ell^j\omega_\ell^i\omega_\ell^jC_\ell^{EE},b_\ell^ib_\ell^j\omega_\ell^i\omega_\ell^jC_\ell^{BB}\big)^{\sf T},\\
  \vec{C}_\ell^{\,\mathrm{dust}} = &~ \big(C_\ell^{E_iB_j,\,\mathrm{dust\,temp.}},C_\ell^{B_iE_j,\,\mathrm{dust\,temp.}}\big)^{\sf T}
\end{align}
containing the power spectra of the observed signal, $\Lambda$CDM CMB prediction, and dust emission template. For every realization of the sky, the covariance matrix $\mathbf{M}_\ell$ is built from the observed signal as detailed in \cite{LiteBIRD:2025mvy}. $\mathbf{A}$, $\mathbf{B}$ and $\mathbf{F}$ are block-diagonal matrices given by
\begin{align}
\nonumber
\mathbf{A}(\boldsymbol{\alpha}) = &~
\mathrm{diag}\left(-\frac{\sin(4\alpha_j)}{\cos(4\alpha_i)+\cos(4\alpha_j)},~\frac{\sin(4\alpha_i)}{\cos(4\alpha_i)+\cos(4\alpha_j)},~1\right)\,,\\
\nonumber
\mathbf{B}(\beta, \boldsymbol{\alpha}) = &~\mathrm{diag}\left(\frac{\sin(4\beta)}{2\cos(2\alpha_i+2\alpha_j)},~ -\frac{\sin(4\beta)}{2\cos(2\alpha_i+2\alpha_j)}\right)\,,\\
\mathbf{F}(A_\mathrm{d}, \boldsymbol{\alpha}) = &~2A_\mathrm{d~}\mathrm{diag}\left(\frac{\cos(2\alpha_i)\cos(2\alpha_j)}{\cos(4\alpha_i)+\cos(4\alpha_j)},~ \frac{\sin(2\alpha_i)\sin(2\alpha_j)}{\cos(4\alpha_i)+\cos(4\alpha_j)}\right)\,.
\end{align}
For computational convenience, we apply the iterative semi-analytical formalism described in \cite{delaHoz:2021vfx,Diego-Palazuelos:2022cnh} to calculate the maximum-likelihood solution of Eq.~(\ref{eq: mk likelihood}) in the small-angle approximation ($|\alpha_i|\ll1$ and $|\beta|\ll1$ ). We bin the input power spectra into uniform bins of $\Delta\ell=10$ width and neglect off-diagonal $\ell\times\ell'$ correlations, expected to be subdominant given the
$2^\circ$ mask apodization scale relative to the bin width.

\begin{figure}[t]
    \centering
    \includegraphics[width=0.8\linewidth]{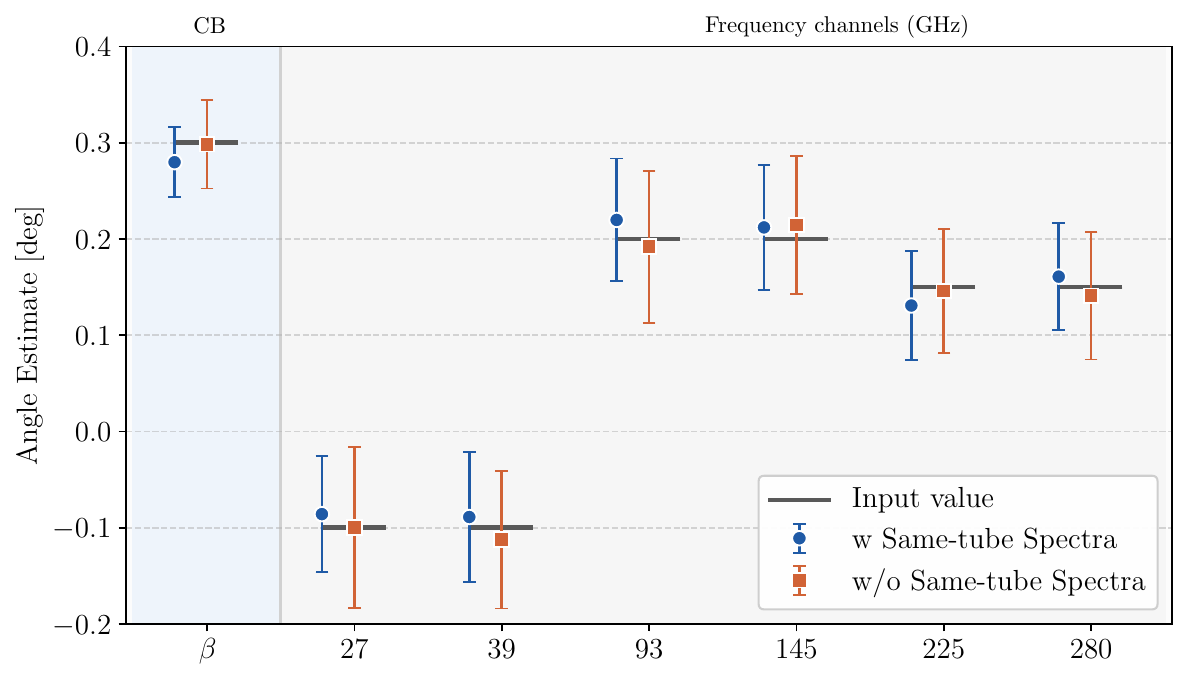}
    \caption{LAT results from the MK estimator applied to 100 LAT simulations with the LAT-I analysis mask and \texttt{d1s1} foregrounds, using SO Baseline noise and $\ell_{\rm max}=3000$. The figure shows the recovered isotropic birefringence angle $\beta$ (left, light-blue panel) and the per-band miscalibration angles $\alpha_i$ at $i=27, 39, 93, 145, 225, 280$\,GHz (right panels). Grey horizontal bars indicate the fiducial values used in the simulations, including $\beta=0.3^\circ$. Two MK configurations are compared: blue circles (\emph{with same-tube spectra}) include all LAT cross-spectra, while orange squares (\emph{without same-tube spectra}) exclude spectra between maps that share the same optical tube.}
    \label{fig:mle_results}
\end{figure}

The simulations used in this section adopt a fiducial birefringence angle
$\beta = 0.3^\circ$ and per-band miscalibration angles $\alpha_i = -0.1^\circ$ at
27 and 39\,GHz, $\alpha_i = 0.2^\circ$ at 93 and 145\,GHz, and $\alpha_i = 0.15^\circ$
at 225 and 280\,GHz. The MK pipeline is applied to 100 LAT simulations using the LAT-I analysis mask, over the baseline LAT multipole range $100 < \ell < 3000$, with results shown in Fig.~\ref{fig:mle_results}. Two configurations are compared: \emph{with same-tube spectra}, which includes all LAT cross-spectra, and \emph{without same-tube spectra}, which excludes cross-spectra between maps
that share the same optical tube.

Excluding all same-tube cross-spectra increases the per-realization scatter but removes the significant mean offset seen when they are included. This comparison does not isolate which same-tube pairs cause the offset; testing a narrower pair selection is left for future work.
The \emph{with same-tube spectra} configuration gives an ensemble mean $\beta=0.280^\circ\pm0.004^\circ$, while the \emph{without same-tube spectra} configuration gives $\beta=0.298^\circ\pm0.004^\circ$; these uncertainties are standard errors of the means across 100 realizations. The corresponding per-realization scatters are $0.036^\circ$ and $0.041^\circ$, respectively. The fitted dust-amplitude means are $\langle A_{\rm d}\rangle=1.019$ with same-tube spectra and $1.021$ without same-tube spectra, close to the unit normalization of the matched dust template. The $0.020^\circ$ offset of the \emph{with-same-tube} mean from the fiducial value is about $5.6$ standard errors in this simulated configuration.\footnote{Note the
consistency with our previously published isotropic CB forecast in Section 3.3 of
Ref.~\cite{so_lat_25}. We obtained $\sigma(\beta)=0.04^\circ$ in this analogous
configuration, excluding spectra from bands in the same optical tube. The previous
forecast used $\ell_{\rm max}=2000$ and goal sensitivity, as opposed to
$\ell_{\rm max}=3000$ and baseline sensitivity used in this work. Overall, the amount
of polarization modes available to constrain CB can be compensated with these choices
to yield a similar constraint.} Removing the same-tube spectra moves the central value closer to the fiducial $\beta=0.3^\circ$ at the price of a $\sim14\%$ increase in the per-realization uncertainty. The \emph{without same-tube spectra} configuration is
adopted as the conservative MK baseline for comparison to the joint LAT+SAT result.

We additionally assess the robustness of the MK pipeline to foreground and bandpass modelling assumptions. The input simulations include bandpass integration and the \texttt{d10s5} foreground model, whereas the dust and synchrotron templates used in the MK likelihood are taken from the \texttt{d9s4} model. We run the estimator both with and without bandpass integration of the foreground templates, and in both cases the recovered $\beta$ remains in statistical agreement with the results reported above, confirming that the pipeline is robust to this level of foreground and bandpass mismatch.

\subsection{Method comparison}
\label{sec:method_comparison}

The two pipelines break the $(\alpha,\beta)$ degeneracy through fundamentally different mechanisms: the joint likelihood imports external calibration priors and, where advantageous, propagates them through the LAT$\times$SAT cross-spectra, while the MK estimator breaks the degeneracy internally using the multi-frequency LAT bands and the distinctness of polarized dust emission.
\begin{figure}
\centering
\includegraphics[width=0.6\columnwidth]{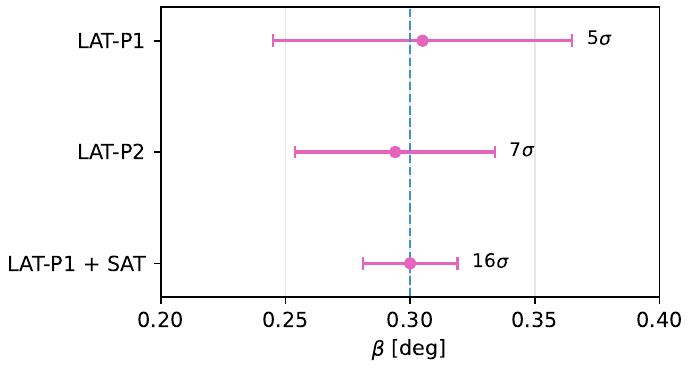}
\caption{Comparison of the recovered cosmic birefringence angle $\beta$ across the two pipelines for a fiducial $\beta=0.3^\circ$ (blue dashed line). The LAT-P1 and LAT-P1 + SAT points use the C3 calibration scenario, SO Goal noise, and the \texttt{d10s5} foreground model with per-frequency amplitude marginalization at 93 and 145\,GHz. The LAT-P2 point uses the conservative six-band MK configuration with SO Baseline noise and no external instrument-angle prior. From top to bottom: ``LAT-P1'' is the LAT-side estimator $\widehat{\beta}_{\rm LAT}$ of Pipeline~1 at the baseline multipole cuts; ``LAT-P2'' is the MK estimator of Pipeline~2; ``LAT-P1 + SAT'' is the inverse-variance-weighted combination of the SAT-side and LAT-side estimators. The number to the right of each error bar gives $\beta_{\rm fid}/\sigma(\beta)$ for the fiducial $\beta_{\rm fid}=0.3^\circ$. The corresponding illustrative uncertainties on $\beta$ are approximately $0.06^\circ$, $0.04^\circ$, and $0.02^\circ$, from top to bottom.}
\label{fig:beta_comparison}
\end{figure}

This distinction translates into a sharp difference in the systematics budgets of the two approaches. For the calibration scenarios considered here, the joint pipeline yields a smaller statistical uncertainty on $\beta$, with that uncertainty scaling with the assumed calibration precision, but its result is conditional on the accuracy of the calibration models of the instrument(s) used as absolute reference. The MK pipeline is internally robust against errors in the external calibration priors, at the price of a larger statistical uncertainty on $\beta$ and a dependence on the assumed dust $EB$ template family, since the degeneracy-breaking power of the method derives directly from the assumed foreground model. The two pipelines therefore probe complementary portions of the calibration and foreground systematics budgets, and their combined use provides a cross-check of any future birefringence detection.

Figure~\ref{fig:beta_comparison} illustrates the selected joint C3/Goal/\texttt{d10s5} and MK Baseline configurations with fiducial $\beta=0.3^\circ$. All three points recover the fiducial value within $1\sigma$ under these assumptions. The ordering of the uncertainties reflects the variance budget of Eq.~\eqref{eq:beta_variance_general}: the transferred LAT-side estimator is overlap-limited at baseline; the MK estimator achieves intermediate precision by trading external-prior information for foreground-assisted degeneracy breaking; and the combined point is dominated by the SAT-side estimate and its tight C3 calibration prior, while the transferred LAT-side estimate provides a check across the LAT multipole range. Expressed as detection significance for the simulated signal, this ordering corresponds to an improvement by a factor of $\sim3$, from $5\sigma$ for the LAT-side estimator, through $7\sigma$ for the MK pipeline, to $16\sigma$ for the combined estimator. For the same input, the combined baseline-sensitivity uncertainties of Table~\ref{tab:mle_results} correspond to a significance rising from $\sim8\sigma$ in C2 to $\sim15\sigma$ in C3, and to $\sim9\sigma$ for the direct estimator in C5. Since $\beta=0.3^\circ$ is a representative value of current hints rather than an established signal, these significances should be read as illustrative only; the robust outputs of the method are the sensitivities $\sigma(\beta)$ and their scaling with the calibration inputs, which remain conditional on the simplifications stated in Secs.~\ref{sec:simulation} and~\ref{sec:sat4lat_likelihood}.

\subsection{Distinguishing between dark-energy and dark-matter birefringence}
\label{sec:de_dm}

The dark-energy and dark-matter birefringence scenarios of \textcolour{blue}{Sec.~\ref{sec:theory}} differ in the scale dependence of $C_\ell^{EB}$, which provides a direct observational handle for discriminating between them. The SO LAT is well suited to this, given its multipole reach to $\ell=3000$.

To illustrate the shape comparison under fixed instrumental-angle assumptions, we perform a Bayesian template comparison. Rather than fitting the underlying particle-physics parameters directly, we compare fixed theoretical $EB$ templates through a single free amplitude parameter $A_{EB}$,
\begin{equation}
\mathcal{D}^{ij}_b(A_{EB}) = \mathcal{D}^{ij,\alpha}_b + A_{EB}\,T^{EB}_b,
\end{equation}
where $\mathcal{D}^{ij,\alpha}_b$ encodes the contribution from the fixed instrumental miscalibration angles and $T^{EB}_b$ is the binned theory template under test. We hold
$\alpha_i$ fixed at its input simulation value rather than marginalizing over it,
since a scale-independent miscalibration produces an $EB$ signal of the same shape as
the constant-$\beta$ dark-energy template, and is therefore already absorbed by the
free amplitude $A_{EB}$ for that hypothesis, as verified at the end of this section.

We consider a scale-independent dark-energy template corresponding to constant birefringence with $\beta=0.3^\circ$, and a family of scale-dependent axion dark-matter templates calculated with \texttt{birefCLASS}~\cite{Nakatsuka:2022epj,Naokawa:2023upt,Murai:2022zur} for masses $m_\phi\in\{0.7,\,0.8,\,0.9,\,1.0,\,1.2,\,1.5,\,6.4\}\times 10^{-28}\,\mathrm{eV}$. These masses are selected from the region of axion-parameter space that remains allowed by current \textit{Planck} data~\cite{Namikawa:2025sft}. For each template, we construct a Gaussian likelihood over the full binned $EB$ data vector, vary only $A_{EB}$, and compute the Bayesian evidence using the \texttt{dynesty}\footnote{\url{https://github.com/joshspeagle/dynesty}}~\cite{dynesty} nested-sampling algorithm. The relative preference between the two scenarios is quantified through the log-Bayes factor
\begin{equation}
\ln B_{\rm DE,DM} \;\equiv\; \ln Z_{\rm DE} - \ln Z_{\rm DM},
\end{equation}
where $Z_{\rm DE}$ and $Z_{\rm DM}$ are the Bayesian evidences of the dark-energy and
dark-matter templates, respectively, each obtained by marginalizing the corresponding
Gaussian likelihood over $A_{EB}$ as described above; with positive values indicating preference for the dark-energy interpretation. The analysis is applied independently to the LAT and SAT simulations with input calibration uncertainties $\sigma(\alpha_{\rm LAT})=0.2^\circ$ and $\sigma(\alpha_{\rm SAT})=0.1^\circ$, \texttt{d1s1} foregrounds and SO baseline noise, with multipole ranges $30\le\ell\le 300$ for the SAT and $100\le\ell\le 3000$ for the LAT. The simulated signal is the dark-energy template with $\beta=0.3^\circ$.

\begin{figure}[t]
\centering
\includegraphics[width=0.6\linewidth]{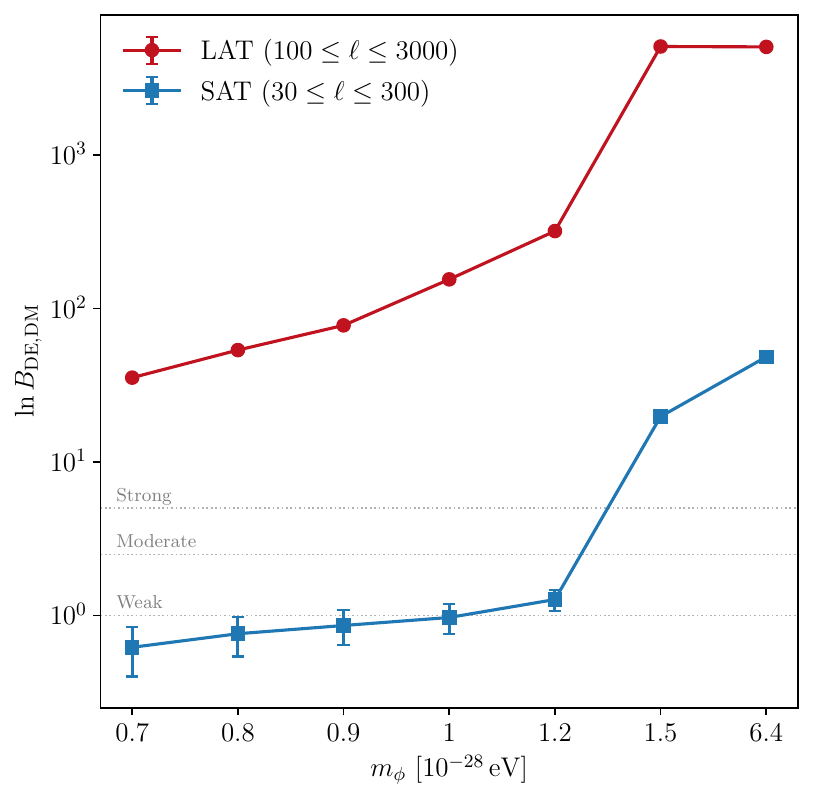}
\caption{Log-Bayes factors $\ln B_{\rm DE,DM}=\ln\mathcal{Z}_{\rm DE}-\ln\mathcal{Z}_{\rm DM}$ discriminating the constant-$\beta$ dark-energy template ($\beta=0.3^\circ$) from axion dark-matter templates of mass $m_\phi$, for the SAT-only ($30\le\ell\le300$) and LAT-only ($100\le\ell\le3000$) analyses. Positive values indicate preference for the dark-energy interpretation; error bars are the nested-sampling evidence errors. Both axes are logarithmic. Dotted lines mark the weak, moderate, and strong thresholds on the Jeffreys scale.}
\label{fig:bayes_factor}
\end{figure}
Figure~\ref{fig:bayes_factor} shows the recovered log-Bayes factors as a function of the axion mass. The two instruments occupy markedly different positions in the discrimination landscape, in direct correspondence with their respective multipole reach. For the SAT, the preference for the dark-energy interpretation is inconclusive ($\ln B_{\rm DE,DM}<1$) at $m_\phi\le 1.0\times 10^{-28}\,\mathrm{eV}$, where the oscillatory features of the dark-matter templates lie predominantly at $\ell>300$, outside the SAT range. The preference rises to weak at $m_\phi=1.2\times 10^{-28}\,\mathrm{eV}$ as the first oscillation ripple enters the SAT range, and becomes decisive for $m_\phi\ge 1.5\times 10^{-28}\,\mathrm{eV}$, with $\ln B_{\rm DE,DM}\simeq20$ and $\simeq49$ at $1.5$ and $6.4\times10^{-28}\,\mathrm{eV}$. For the LAT, the picture is qualitatively different: with access to $\ell=3000$, it captures the full oscillatory pattern of every mass in the scan and yields decisive discrimination across the entire range $m_\phi\in[0.7,6.4]\times 10^{-28}\,\mathrm{eV}$, with log-Bayes factors ranging from $\ln B_{\rm DE,DM}\simeq36$ at the lightest mass to several thousand at the heaviest, orders of magnitude larger than the SAT for the lighter masses. The slight decrease at $m_\phi=6.4\times 10^{-28}\,\mathrm{eV}$ relative to $1.5\times 10^{-28}\,\mathrm{eV}$ reflects the partial averaging of the most rapidly oscillating templates within the $\Delta\ell=10$ bandpowers. This averaging could in principle be mitigated with a variable bin width, narrower at high $\ell$ where the templates oscillate most rapidly and wider at low $\ell$, which would better resolve the oscillatory structure and recover part of the lost discrimination power, at the cost of a noisier per-bin covariance; since the LAT is already decisive across the full mass range, we do not pursue this optimization here, but it may be valuable for a real analysis aiming to pin down the axion mass rather than merely establish the dark-matter interpretation.

These Bayes factors are conditional on fixing the instrumental angles and the template set; marginalizing the angles with calibration priors could change the model preference and is not included here. They are also conditional on the absence of scale-dependent residual systematics, against which they have not been stress-tested.

Crucially, this comparison is driven by the multipole \emph{shape} of $C_\ell^{EB}$ rather than its overall amplitude: since $A_{EB}$ is marginalized over for every template, the preference for the constant-$\beta$ model arises from the inability of the oscillatory dark-matter templates to reproduce the multipole structure of the fiducial $EB$ spectrum. The Bayes factor therefore compares the shapes of the templates, but a preference for the constant-rotation template alone cannot establish a cosmological origin: an uncalibrated instrumental angle offset can produce the same shape. The similar Bayes factors obtained from calibrated and uncalibrated LAT maps illustrate this degeneracy; absolute-angle calibration is required to attribute the constant component to $\beta$.

\section{Discussion and conclusions}
\label{sec:discussion}

We use SO LAT and SAT map-level simulations to study how isotropic birefringence inference depends on absolute-angle calibration and the multipole range shared by the instruments. Our primary method is a joint LAT+SAT likelihood that transfers calibration through their common CMB modes; we also test the Minami--Komatsu foreground-based estimator as a complementary check, conditional on the assumed Galactic dust $EB$ model.

A central result of this study is that polarization-angle calibration remains the
dominant limiting systematic for isotropic birefringence measurements with the SO LAT,
within the systematics modeled here, and that how the available calibrations are combined matters as much as how good they are. The general variance budget of Eqs.~\eqref{eq:beta_variance_general}--\eqref{eq:beta_comb} shows that transferring a calibration from one instrument to the other pays only when the reference calibration, combined in quadrature with the relative-angle uncertainty $\sigma(\Delta\alpha)$, is smaller than the receiving instrument's own (Eq.~\eqref{eq:transfer_condition}). When the SAT is better calibrated (C2, C3), progressive improvements in the SAT calibration translate into substantial gains in sensitivity: the combined uncertainty improves from $\sigma(\beta)\simeq0.04^\circ$ in C2 to $\simeq0.02^\circ$ in C3. When the SAT provides the stronger absolute reference, transfer chiefly enables angle-calibrated LAT $EB$ measurements over a wider multipole range; it does not make the SAT and LAT estimates independent measurements of the absolute birefringence angle. When the two calibrations are comparable (C4) or the LAT is better calibrated (C5), transferring the SAT calibration to the LAT is suboptimal, and anchoring each instrument on its own prior improves $\sigma(\beta_{\rm LAT})$ from $0.11^\circ$ to $0.07^\circ$ and $0.04^\circ$, and $\sigma(\beta)$ from $0.06^\circ$ to $0.05^\circ$ and $0.03^\circ$, respectively. Because the transferred LAT-side estimator shares the absolute prior with the SAT-side one, the combination cannot average that prior down, and the combined values for $\sigma(\beta)$ quoted for transferred configurations are lower bounds.

Under the baseline multipole cuts, the transferred LAT-side estimator operates in the overlap-limited regime, where the bottleneck is the LAT--SAT relative-angle determination rather than the reference prior itself. The primary lever for improvement is therefore the SAT high-$\ell$ reach, with $\sigma(\Delta\alpha)$ improving by a factor of $\sim3$ when $\ell_{\rm max}^{\rm SAT}$ is extended from 300 to 600. A larger overlap would also turn the relative angle into a useful second anchor for comparable calibrations, and would make the reverse transfer of a better LAT calibration to the SAT worthwhile. Comparing the SO Baseline and Goal noise configurations shows a modest but meaningful improvement in $\sigma(\beta_{\rm LAT})$ from $0.09^\circ$ to $0.06^\circ$, driven by the corresponding reduction in $\sigma(\Delta\alpha)$, but substantially smaller than the gain from extending the SAT multipole reach: for the transferred LAT-side estimator, SAT high-$\ell$ calibration and transfer-function control are worth more than raw sensitivity.

Polarized Galactic foregrounds are handled in the joint pipeline by per-frequency amplitude marginalization, under which the marginalized posteriors on $\beta$ remain unbiased across both foreground models and both Galactic masks. The MK estimator provides a complementary cross-check: its conservative \emph{without same-tube spectra} configuration yields an ensemble mean $\beta=0.298^\circ\pm0.004^\circ$ across 100 realizations, with per-realization scatter $0.041^\circ$, in statistical agreement with the joint LAT+SAT recovery and with the fiducial value, and is stable against foreground template mismatch and bandpass integration.

The two pipelines are complementary in their systematics budgets (Sec.~\ref{sec:method_comparison}), and their combined use provides a cross-check of any future birefringence detection. For a simulated $\beta=0.3^\circ$, taken as a representative value of current hints rather than an established signal, the detection significance spans a factor of $\sim3$ across the configurations examined, from $5\sigma$ for the LAT-side estimator alone to $16\sigma$ for the combined estimator under the most optimistic C3/Goal/\texttt{d10s5} configuration. These significances are illustrative; the central output of the method is the scaling of $\sigma(\beta)$ with the calibration inputs and its traceability to the variance budget of Eq.~\eqref{eq:beta_variance_general}.

Finally, our fixed-template comparison shows how the LAT multipole range can help distinguish the $C_\ell^{EB}$ shapes of constant-rotation and axion dark-matter interpretations under the stated instrumental-angle assumptions. The LAT fixed-template Bayes factors favor the constant-rotation template across the mass range $m_\phi\in[0.7,6.4]\times10^{-28}\,\mathrm{eV}$ explored here, while the SAT has weaker shape discrimination at the lower masses. This preference alone does not establish a cosmological origin for a constant rotation.

\subsection*{Future extensions}

Several extensions of the present work would further improve the SO sensitivity to isotropic cosmic birefringence. Including the $TB$ spectrum in addition to $EB$ is expected to tighten the joint constraint and improve the disentanglement of cosmological and instrumental rotations. Internal LAT-based delensing would reduce the lensing-induced $B$-mode variance and improve $\sigma(\beta)$ in the noise-limited regime. On the foreground side, more sophisticated dust models incorporating filamentary morphology, line-of-sight decorrelation, and frequency-dependent polarization-angle variations should be tested. Future iterations of the analysis should also incorporate anisotropic noise, frequency- and detector-dependent beam mismatch, correlated SAT--LAT calibration uncertainties, a full bandpower covariance including the correlations between likelihood blocks and between the LAT- and SAT-side estimators, and realistic time-domain systematics propagated map-level through the simulation pipeline. The optimally weighted estimator of Eq.~\eqref{eq:beta_variance_twosided} and the reverse LAT-to-SAT transfer should be run through the full pipeline, in particular at extended $\ell_{\rm max}^{\rm SAT}$ where they are expected to pay off. A further extension is to quantify the discrimination power of the consistency test described in Sec.~\ref{sec:sat4lat_regimes}: injecting a deliberately offset angle into the simulations and characterizing the resulting distribution of the test statistic would establish, for a given pair of calibration scenarios, the smallest genuine miscalibration this framework could detect as an internal tension rather than absorb into the combined estimate. These extensions are natural next steps as the SO acquires its first years of polarization data and the calibration program matures toward the precision targets explored in this work.

\appendix

\section{Map-based LAT--SAT cross-calibration}
\label{sec:app_map_calib}

The baseline estimator of $\Delta\alpha$ (Sec.~\ref{sec:sat4lat_deltaalpha}) operates on bandpower-compressed data. Here we describe a complementary map-level estimator that operates directly on the LAT and SAT polarization maps in pixel space, before any harmonic compression. Such an approach retains spatial information about position-dependent miscalibrations that is averaged over in rotationally-symmetric bandpowers, and may be advantageous in the presence of anisotropic noise or spatially varying beam effects.

\subsection*{Map-level estimator}

Defining the complex polarization field $P(\hat n)=Q(\hat n)+iU(\hat n)$, the observed LAT and SAT maps satisfy
\begin{equation}
P_{L}^{\mathrm{obs}}(\hat n)
= e^{2i(\alpha_{L}+\beta)}P^{\mathrm{true}}(\hat n)+n_{L}(\hat n),
\qquad
P_{S}^{\mathrm{obs}}(\hat n)
= e^{2i(\alpha_{S}+\beta)}P^{\mathrm{true}}(\hat n)+n_{S}(\hat n),
\label{eq:P_LS_obs}
\end{equation}
so that the cosmic birefringence angle cancels in the cross-product,
\begin{equation}
P_{L}^{\mathrm{obs}}(\hat n)\,\bigl[P_{S}^{\mathrm{obs}}(\hat n)\bigr]^{\ast}
\;=\; e^{2i\Delta\alpha}\,\bigl|P^{\mathrm{true}}(\hat n)\bigr|^{2}
\;+\; \mathcal{N}(\hat n),
\label{eq:cross_product}
\end{equation}
where $\mathcal{N}=e^{2i(\alpha_L+\beta)}P^{\mathrm{true}}n_S^{\ast}+e^{-2i(\alpha_S+\beta)}n_LP^{\mathrm{true}\ast}+n_Ln_S^{\ast}$ collects the noise terms. Since the LAT and SAT noise realizations are independent of each other and of the sky, $\langle\mathcal{N}\rangle=0$ and the noise does not bias $\widehat{\Delta\alpha}$ at first order; it contributes only to its variance, which we quantify directly from simulations below. We therefore drop $\mathcal{N}$ in what follows. We note also that the weights $w_p$ in Eq.~\eqref{eq:AB_map} need not be diagonal: replacing the input maps by their inverse-variance ($C^{-1}$) filtered counterparts gives the optimal weighting in the presence of anisotropic noise and masking, at the cost of a matrix inversion, and would further suppress the noise contribution to the variance.
The phase of the weighted pixel sum $Z=\sum_{p}w_{p}P_{L,p}P_{S,p}^{\ast}$ therefore provides an estimator of the LAT$-$SAT relative miscalibration,
\begin{equation}
\widehat{\Delta\alpha} \;=\; \tfrac{1}{2}\,\arctan2\,(B,\,A),
\label{eq:dalpha-map}
\end{equation}
with
\begin{equation}
A=\sum_{p}w_{p}\bigl(Q_{L,p}Q_{S,p}+U_{L,p}U_{S,p}\bigr),
\qquad
B=\sum_{p}w_{p}\bigl(U_{L,p}Q_{S,p}-Q_{L,p}U_{S,p}\bigr).
\label{eq:AB_map}
\end{equation}

\subsection*{Validation on SO simulations}

We have implemented and validated this estimator on the SO map-level simulations used in the main analysis, with input calibration uncertainties $\sigma(\alpha_{\rm LAT})=0.2^\circ$ and $\sigma(\alpha_{\rm SAT})=0.1^\circ$ and without simulated cosmic birefringence and with the \texttt{d1s1} sky model for polarized Galactic foregrounds. Three preprocessing steps are applied before evaluating Eq.~\eqref{eq:dalpha-map}.

First, both LAT and SAT maps are high-pass filtered to remove modes with $\ell< 100$, which is the lower edge of the multipole range over which LAT $EB$ bandpowers enter the baseline likelihood; restricting the map-level estimator to the same range ensures that LAT and SAT are compared on the band of scales that is common to both analyses. Second, the two maps are convolved to a common Gaussian beam $\theta_{\mathrm{c}}$, which equalizes their angular resolution and is the operation that allows the factorization $P_{L}^{\mathrm{obs}}\propto e^{2i\Delta\alpha}P_{S}^{\mathrm{obs}}$ implicit in Eq.~\eqref{eq:dalpha-map} to hold to good approximation. Third, the binary LAT$\cap$SAT overlap mask is apodized with a C2 taper of extent $\theta_{\mathrm{apo}}=2^{\circ}$, using the \texttt{NaMaster} convention, to suppress $E\!\to\!B$ leakage from the mask boundary into the spin-2 smoothing; the pixel weight is taken to be the apodized mask, $w_{p}=M_{\mathrm{apo},p}$.

\begin{figure}[t]
\centering
\includegraphics[width=0.6\linewidth]{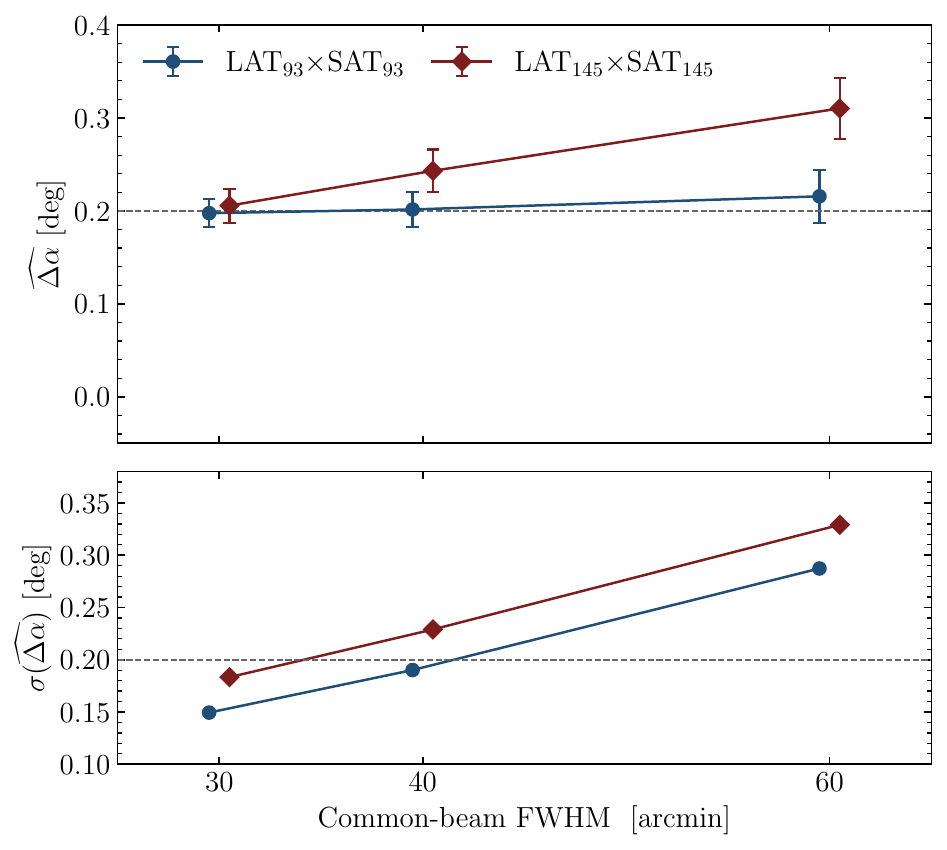}
\caption{Validation of the map-level relative-calibration estimator on SO map-level simulations ($\sigma(\alpha_{\rm LAT})=0.2^\circ$, $\sigma(\alpha_{\rm SAT})=0.1^\circ$, \texttt{d1s1} foregrounds). \emph{Top:} Monte Carlo mean of $\widehat{\Delta\alpha}$ [Eq.~\eqref{eq:dalpha-map}] as a function of the common-beam FWHM $\theta_{\mathrm{c}}$, for same-frequency, LAT$\times$SAT pairs at 93 and 145\,GHz; error bars are the standard error computed with $N_{\mathrm{sim}}=100$. \emph{Bottom:} per-realization scatter $\sigma(\widehat{\Delta\alpha})$ for the same pairs. The dashed grey line shows the mean and $1\sigma$ values of the input miscalibration.}
\label{fig:maplevel-fwhm}
\end{figure}

Figure~\ref{fig:maplevel-fwhm} shows the Monte Carlo mean and standard deviation of $\widehat{\Delta\alpha}$ over $N_{\mathrm{sim}}=100$ realizations, for the four same-frequency, same-split-index LAT$\times$SAT pairs, at three values of the common beam, $\theta_{\mathrm{c}}\in\{30',\,40',\,60'\}$. The top panel shows that the recovered $\widehat{\Delta\alpha}$ is consistent across the four pairs at $\theta_{\mathrm{c}}=30'$ and traces the fiducial LAT--SAT miscalibration. The bottom panel shows that the per-realization scatter $\sigma(\widehat{\Delta\alpha})$ grows approximately linearly with $\theta_{\mathrm{c}}$, consistent with the mode-count expectation $\sigma\propto\theta_{\mathrm{FWHM}}$ since the number of independent polarization modes contributing to the sum scales as $\theta_{\mathrm{c}}^{-2}$. At $\theta_{\mathrm{c}}=30'$ the per-pair scatter is $\sigma(\widehat{\Delta\alpha})\approx 0.15^{\circ}$, comparable to but not improving on the $0.1^\circ$ SAT absolute-calibration prior assumed in these simulations.

Cross-frequency LAT$\times$SAT pairs are excluded from this validation because the underlying polarized sky differs between 93 and 145\,GHz under the \texttt{d1s1} model (dust and synchrotron have different SEDs), violating the factorization in Eq.~\eqref{eq:cross_product} and producing a foreground-induced bias in $\widehat{\Delta\alpha}$. A joint combination with the cross-spectrum estimator, and the treatment of the instrumental effects not modelled here (Sec.~\ref{sec:pol_angle}), are left to future work.

\section{Extension to $TB$ cross-spectra}
\label{sec:ext_TB}
The $TB$ cross-spectrum provides an independent handle on $\Delta\alpha$ that complements the $EB$-based estimator. It can be shown that the antisymmetric combination of the LAT$\times$SAT $TB$ cross-spectra satisfies
\begin{equation}
C_\ell^{T_LB_S} - C_\ell^{T_SB_L} = -2\,C_\ell^{TE}\cos(\theta_L+\theta_S)\sin(\Delta\alpha),
\end{equation}
which is independent of $\beta$ at leading order in the small-angle approximation, and constrains $\Delta\alpha$ in a manner complementary to Eq.~\eqref{eq:antisym_combo}. Unlike the $EB$-based antisymmetric combination, where $\beta$ cancels identically, the $TB$ estimator retains residual $\beta$ dependence at higher order; however, for the small rotation angles of interest this correction is negligible. Using $TB$ alone yields a weaker constraint on $\Delta\alpha$ than $EB$ alone, by roughly a factor of 1.7 at both frequencies, reflecting the lower signal-to-noise of the $TB$ spectrum. Combining $TB$ and $EB$ jointly improves $\sigma(\Delta\alpha)$ by $\sim\!14\%$ over the $EB$-only result, with consistent gains at both 93 and 145\,GHz. While the improvement is modest, the $TB$ block provides a genuinely independent cross-check of the $EB$-derived $\Delta\alpha$, since it relies on temperature rather than $E$-mode polarization and is therefore subject to different systematic contamination. We leave a full assessment of the $TB$ contribution across all calibration scenarios and its propagation into $\sigma(\beta)$ for future work.

\acknowledgments

This work was funded by grants from the Simons Foundation [MPS-Observatory-00457687, B.K.] and Simons Foundation International [SFI-MPS-Observatory-00007127, B.K.]. SO operates in the Parque Astronómico Atacama in northern Chile under the auspices of the Agencia Nacional de Investigación y Desarrollo (ANID). We thank the Republic of Chile for hosting SO in the northern Atacama, and the local indigenous Licanantay communities, whom we follow in observing and learning from the night sky.  
This research used resources of the National Energy Research Scientific Computing Center (NERSC), a Department of Energy User Facility (HEP project mp107 2023-2026). 
This work was supported by the U.S. National Science Foundation (Award Number: 2153201). 
Some of the software developed to produce this analysis can be found publicly available on the Simons Observatory GitHub organization (https://github.com/simonsobs). The code used to generate the simulations and perform the analysis in this study is publicly available on GitHub at \url{https://github.com/antolonappan/cobi}.
The authors thank Fumihiro Naokawa for providing the scale-dependent axion dark-matter theoretical CMB spectra. 

P.D.-P. acknowledges support by the Deutsche Forschungsgemeinschaft (DFG, German Research Foundation) under Germany's Excellence Strategy - EXC-2094/2 - 390783311.
C.H.-C. acknowledges support from Agencia Nacional de
Investigaci\'on y Desarrollo (ANID) FONDECYT regular 1262583 and BASAL CATA, FB210003.
C.B. acknowledges partial support from the Italian Space Agency
LiteBIRD Project (ASI Grants No.~2020-9-HH.0 and
2016-24-H.1-2018), the Italian Space Agency Euclid Project,
the InDark and LiteBIRD Initiative of the National Institute
for Nuclear Physics, Project SPACE-IT-UP by the Italian Space
Agency and Ministry of University and Research
(Contract No.~2024-5-E.0), the RadioForegroundsPlus Project
(HORIZON-CL4-2023-SPACE-01, Grant Agreement No.~101135036),
and the CMB-Inflate project funded by the European Union's
Horizon 2020 Research and Innovation Staff Exchange under
the Marie Sk{\l}odowska-Curie Grant Agreement No.~101007633.
Y.C. acknowledges support from JSPS KAKENHI
Grant No.~JP24K00667.
J.E. and A.L. acknowledge funding from the \textsc{SciPol}
project (\href{https://scipol.in2p3.fr}{scipol.in2p3.fr}),
supported by the European Research Council (ERC) under
the European Union's Horizon 2020 research and innovation
program (Grant Agreement No.~101044073, PI: J. Errard).
T.N. acknowledges support from JSPS KAKENHI
Grant Nos.~JP25K00996 and JP26H00405, and from
JST EXPERT-J, Japan, Grant No.~JPMJEX2508.
The Kavli IPMU is supported by the World Premier
International Research Center Initiative (WPI Initiative),
MEXT, Japan.
F.N. acknowledges funding from the European Union
(ERC, POLOCALC, 101096035).
K.W. is supported by the Science and Technology Facilities
Council (STFC) under grants ST/X006344/1 and UKRI1164,
and acknowledges support from the Beecroft Trust.
\bibliographystyle{jcap/JHEP}
\bibliography{misc/reference}
\end{document}